\documentclass[%
 aip,
amsmath,amssymb,
preprint,
]{revtex4-1}

\usepackage{graphicx}
\usepackage{dcolumn}
\usepackage{bm}

\usepackage[utf8]{inputenc}
\usepackage[T1]{fontenc}
\usepackage{mathptmx}
\usepackage{etoolbox}

\usepackage{mathtools, amsthm}
\usepackage{array}
\usepackage{multirow}
\usepackage{tikz} 
 \usetikzlibrary{arrows,backgrounds,patterns}
  \usetikzlibrary{decorations.markings}
  \usetikzlibrary{circuits.logic.US} 
  \usetikzlibrary{external}
\definecolor{mygreen}{rgb}{0,0.6,0}
\definecolor{darkgreen}{rgb}{0.1,0.7,0.3}
\definecolor{mygray}{rgb}{0.5,0.5,0.5}
\definecolor{mymauve}{rgb}{0.58,0,0.82}
\definecolor{lightgray}{rgb}{0.5,0.5,0.5}

\newcommand{\OR}{\texttt{OR}}
\newcommand{\XOR}{\texttt{XOR}}

\newcommand{\R}{\mathbb{R}}

\newtheorem{prop}{Proposition}

\makeatletter
\def\@email#1#2{%
 \endgroup
 \patchcmd{\titleblock@produce}
  {\frontmatter@RRAPformat}
  {\frontmatter@RRAPformat{\produce@RRAP{*#1\href{mailto:#2}{#2}}}\frontmatter@RRAPformat}
  {}{}
}%
\makeatother
\begin{document}

\preprint{AIP/123-QED}


\title[Chaos in a switching network]{Computer-aided analysis of high-dimensional Glass networks:\\ periodicity, chaos, and bifurcations in a ring circuit}
\author{I. Belgacem}
 \affiliation{Mathematics \& Statistics, University of Victoria, Canada.}
\author{R. Edwards}
 \affiliation{Mathematics \& Statistics, University of Victoria, Canada.}
\author{E. Farcot}%
 \email{Etienne.Farcot@nottingham.ac.uk}
\affiliation{ School of Mathematical Sciences, University of Nottingham, UK.
}%

\date{\today}

\begin{abstract}
Glass networks model systems of variables that interact via sharp switching. A body of theory has been developed over several decades that, in principle, allows rigorous proof of dynamical properties in high dimensions that is not normally feasible in nonlinear dynamical systems. Previous work has, however, used examples of dimension no higher than $6$ to illustrate the methods. Here we show that the same tools can be applied in dimensions at least as high as $20$. An important application of Glass networks is to a recently-proposed design of a True Random Number Generator that is based on an intrinsically chaotic electronic circuit. In order for analysis to be meaningful for the application, the dimension must be at least $20$. Bifurcation diagrams show what appear to be periodic and chaotic bands. Here we demonstrate that the analytic tools for Glass networks can be used to rigorously show where periodic orbits are lost, and the types of bifurcations that occur there. The main tools are linear algebra and the stability theory of Poincar\'e maps. All main steps can be automated, and we provide computer code. The methods reviewed here have the potential for many other applications involving sharply switching interactions, such as artificial neural networks.
\end{abstract}

\maketitle

\begin{quotation}
There are few examples where a bifurcation to chaos can be characterized using analytical methods. There are much fewer still where this can be conducted in a high- (e.g. 20-) dimensional state space. Here we demonstrate how this can be achieved for ``Glass network'' models, a class of piecewise-linear differential equations used to model gene, neural and electronic circuits. We provide general background on the methods, and then focus on a specific example which relies on chaos to serve as random number generator. We provide software allowing the reader to conduct similar analyses on any Glass network.
\end{quotation}

\ \\
\section{Introduction}

There are few examples of high-dimensional nonlinear dynamical systems in which it is possible to demonstrate rigorously the existence or stability of periodic orbits. Transitions to chaos are more difficult still. Glass networks, which are a class of piecewise-linear switching systems, are an exception. In this class of highly nonlinear systems it is possible to obtain analytic results about periodic orbits and bifurcations in which they are lost (or gained), and to do it in quite a high-dimensional context. An extensive body of analysis techniques has been developed over several decades for $N$-dimensional systems (arbitrary $N$), but the examples considered are always relatively low-dimensional ($N\le 6$), apart from numerical studies. Here, in the context of an important applied problem, we show by actually carrying out the analysis (with the aid of a computer, to be sure) that it is practicable to apply these methods in dimensions as high as 20, with periodic orbits with hundreds of transition steps. This is quite remarkable, and the methods deserve to be better known. 

The applications of Glass networks are to simple switching systems, including gene networks (or at least simplified qualitative models of gene networks)~\cite{glass1973logical,glass1975combinatorial,dejong2004qualitative,glass2018hybrid}, neural networks~\cite{glass1979structure,lewis1992nonlinear,edwards1999parkinsonian,edwards2003synchronization}, and free-running electronic circuits, like those used as True Random Number Generators (TRNGs)~\cite{rosin2013ultrafast,farcot2019chaos,luo2020high}, but the diversity of these applications strongly suggests that there will be others. In this last application at least, chaotic dynamics are desirable, so chaotic attractors are of interest, in addition to steady states and periodic orbits. In the context of neural networks, what is likely to be relevant is the co-existence of multiple attractors of any type, as well as transitions between them, in a large dimensional state space. It is worth noting that, after a simple coordinate change, Glass networks include Hopfield neural networks as a special case~\cite{lewis1991steady}. The latter are still relevant in the context of deep learning in general and transformers in particular \cite{krotov2023new,ramsauer2020hopfield}. 

As a working example, we focus on an electronic circuit design recently proposed as a robust TRNG. The steps we follow, however, are generic and could be applied to any Glass network. In a previous paper~\cite{farcot2019chaos}, Scott Best of Rambus, Inc. (California), proposed this new design, 
which as well as having the randomizing effect of thermal noise in the circuit, has an inherently chaotic underlying dynamics, making it more robust to frequency-injection attack than TRNGs based on simple oscillators. It is composed of a ring of $n$ units ($n$ odd, but $\ge 5$), each of which has four standard logic gates, with feedback and feedforward between neighbouring units. In this earlier paper, we analyzed the structure of the dynamics in the noise-free situation.
The behaviour could be determined rigorously for simplified models of this circuit, in which two or three of the variables are taken to operate on an infinitely-fast time scale, resulting in models with $2n$ and $n$ variables, respectively. For the full $4n$-dimensional model of the circuit, however, we only proved existence of chaos through Lyapunov Exponents estimated from output of numerical simulations, although we suggested that chaos might arise as the inverters and OR gate go from infinitely fast to very fast, as a result of the oscillations that arise around the sliding mode in the $n$-dimensional version of the model.

Now we improve on the earlier work by investigating the bifurcations that occur in the $4n$-dimensional circuit with $n=5$ when decay rates of the gates are taken to be equal and units are taken to be identical. These are not quite physically accurate approximations but similar bifurcation diagrams are found with wide ranges of parameter values, so this case should be typical, and we have strong analytic tools for the equal decay rate case. Poincar\'e maps can be calculated exactly for any cycle of states in the state transition diagram, and conditions for existence of a stable periodic orbit for such a cycle can be checked by exact calculation. Of course, the dimension is high ($20$) and the number of steps in the relevant cycles tends to be large ($\approx 100$ or more), so the calculations are automated and use extended precision computations. A numerically-computed bifurcation diagram (as one of the parameters is varied) shows a complicated series of transitions between periodic windows and (apparently) fully chaotic intervals of parameter values, with other bifurcation points in both regimes. The analytic methods can determine the nature of these bifurcations rigorously.

This is stronger evidence for chaos in this circuit than was provided only numerically in the previous paper, since at a transition to chaos, the bifurcation is of a saddle-node type, where a stable and unstable periodic orbit collide and vanish, so locally at least there is no periodic orbit after the bifurcation. It is still possible in principle that after the bifurcation point another stable periodic orbit exists at some distance in phase space, but the numerical evidence makes this unlikely.

The fact that we can identify these bifurcation points, determine their type and the nature of the limit cycles on either side (if any) in $20$ dimensions, all by rigorous, though computer-aided, analysis seems to the authors worth highlighting. The application to TRNGs is an important one, and 
the methods described here may be useful in other applications.

We begin with the necessary background on the ring circuit and the differential equations describing it (in various forms) in Section~\ref{sec:BestCircuit}, on the basic theory for Glass networks in Section~\ref{sec:glassnets}, and on bifurcations of cycles in Glass networks in Section~\ref{sec:bifurcations}. Then in Section~\ref{sec:10d}, we ease our way into analysis of the ring circuit by way of a simplified ($2n=10$-dimensional) model, where we can illustrate the calculations needed to prove existence and stability of a periodic orbit, and show that in this model it is not lost through bifurcation.
Then in Section~\ref{sec:4n} we analyze the full $4n$-dimensional model with $n=5$ for a range of parameter values, identifying rigorously the nature of several of the many bifurcations that occur, including a transition from a periodic regime to a chaotic regime. Finally, we conclude with a discussion in Section~\ref{sec:discussion}.
 
\section{The ring circuit}\label{sec:BestCircuit}
\begin{figure}[h]
\begin{center}
\begin{tikzpicture}[circuit logic US, every circuit symbol/.style={thick}, scale=.8]
\def \lw {1.5}
\coordinate (corner) at (0,0);
\draw[line width=\lw,color=black!50,fill=black!25!white, rounded corners=4] (corner) rectangle ++(7.5,-3.5);
\draw[line width=.5*\lw,color=black!20,fill=black!5!white, rounded corners=4] (corner) ++(4mm,-2mm) rectangle ++(2.8,-3);

\node[xor gate, point right, inputs={nn},fill=white] (xi) at ([xshift=20mm, yshift=-10mm] corner) {}; %
\node[not gate, point right,fill=white] (yi) at ([xshift=40mm, yshift=-10mm] corner) {}; %
\node[not gate, point right,fill=white] (zi) at ([xshift=60mm, yshift=-10mm] corner) {}; %
\node[or gate, point right,fill=white] (ui) at ([xshift=15mm, yshift=-25mm] corner) {}; 
\draw[line width=\lw, rounded corners] (xi.output) node[above] {$x_i$} --  (yi.input);
\draw[line width=\lw, rounded corners] (yi.output) node[above] {$y_i$} --  (zi.input);
\draw[line width=\lw, rounded corners] (ui.output) node[above] {$u_i$} -- ++(right:7mm) -- ++(up:8mm) -- ++(left:16mm) -- ++(up:6mm) -- (xi.input 2);%
\draw[line width=\lw, rounded corners] (zi.output) -- ++(right:5mm) -- ++(down:20mm) -- ++(left:65mm) -- ++(up:4mm) -- (ui.input 2);
\draw[line width=\lw, rounded corners, dotted, ->] (zi.output) -- ++(right:15mm) -- ++(down:8mm)  node[xshift=1mm,yshift=-3mm] {($u_{i-1})$};

\draw[line width=\lw] (zi.output) --   +(right:13mm);
\draw[line width=\lw, rounded corners, dotted, ->] (zi.output) node[above]{$z_i$} --   +(right:25mm) node[right] {$(x_{i+1})$};

\draw[line width=\lw, rounded corners] ([xshift=-10mm, yshift=-9mm] corner) node[above] {$z_{i-1}$} -- (xi.input 1);
\draw[line width=\lw, dotted] ([xshift=-15mm, yshift=-9mm] corner) -- +(1,0);
\draw[line width=\lw, rounded corners] ([xshift=-12mm, yshift=-27mm] corner) -- ++(up:3mm)  node[above,xshift=2mm] {$z_{i+1}$} -- (ui.input 1);
\draw[line width=\lw, rounded corners, dotted] ([xshift=-12mm, yshift=-31mm] corner) -- ++(up:6mm)  ;

\end{tikzpicture} 
\end{center}
\caption{\label{fig:circuit} The basic unit (dark shaded area) receives the voltages $z_{i\pm 1}$ from the previous and next units as inputs. Each unit comprises the Boolean function $f(a,b,c)= a \,\oplus\, (b\,\vee\, c)$ encoded by means of an \OR\ and an \XOR\ gate (light shaded area), as well as two inverters. The overall structure is periodic: $i\in \{1,...,n\}$, $i$ is considered modulo $n$, and $n$ is odd.}
\end{figure}
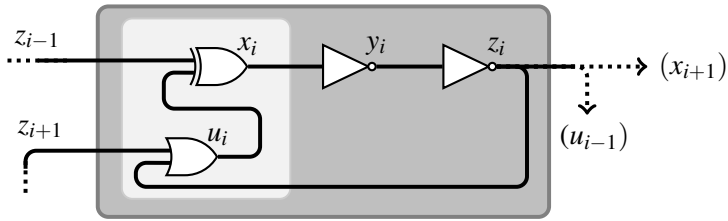

The ring circuit \cite{farcot2019chaos, luo2020high} is depicted in Figure~\ref{fig:circuit}.

According to the scheme in Figure~\ref{fig:circuit}, every unit can be described by the following system of Ordinary Differential Equations (ODEs):
\begin{equation}\label{eq:4n}
\begin{array}{lcl}
\displaystyle \frac{dx_i}{dt} & = & \kappa_{x_i} \left( s^+(z_{i-1})s^-(u_i) + s^-(z_{i-1})s^+(u_i)\right) - \gamma_{x_i}x_i  \\[3mm]
\displaystyle \frac{dy_i}{dt} & = & \kappa_{y_i}s^-(x_i) - \gamma_{y_i}y_i\\[3mm]
\displaystyle \frac{dz_i}{dt} & = & \kappa_{z_i}s^-(y_i) - \gamma_{z_i}z_i\\[3mm]
\displaystyle \frac{du_i}{dt} & = & \kappa_{u_i}\left(1 - s^-(z_{i})s^-(z_{i+1})\right) - \gamma_{u_i}u_i\,.
\end{array}
\end{equation}
The $s^+(\cdot)\in\{0,1\}$ terms denote the Heaviside function with a threshold, denoted $\theta_{x_1}$, etc., whilst $s^-=1-s^+$. Since we have only one threshold for each variable, it will be convenient to translate the variables so that thresholds fall at the origin. In this context, we use a single letter notation: $v=(v_i) \in \mathbb{R}^{4n}$ is a vector of voltages, satisfying the above equations with the substitution $(v_{4i-3},v_{4i-2},v_{4i-1},v_{4i})=(x_i-\theta_{x_i},y_i-\theta_{y_i},z_i-\theta_{z_i},u_i-\theta_{u_i})$ and $v$ subscripts being understood modulo $4n$. For the most part we will also assume all thresholds identical: $\theta_{x_i}=\theta_{y_i}=\theta_{z_i}=\theta_{u_i}\equiv \theta$ for all $i=1,\ldots, n$.

\bigskip 

As a simplified, lower-dimensional model, we will also consider the case where the dynamics of the two inverters are fast compared to other components, i.e. $\kappa_{x_i}$, $\kappa_{u_i}\ll \max\{\kappa_{y_i}, \kappa_{z_i}\}$, and  
$\gamma_{x_i}$, $\gamma_{u_i}\ll \max\{\gamma_{y_i}, \gamma_{z_i}\}$, $i\in\{1,\ldots , n\}$ and the small quantities are set to zero. This leads to a $2n$ dimensional model:
\begin{equation}\label{eq:2n2}
\begin{array}{lcl}
\displaystyle \frac{dx_i}{dt} & = & \kappa_{x_i} \left( s^+(x_{i-1})s^-(u_i) + s^-(x_{i-1})s^+(u_i)\right)  - \gamma_{x_i}x_i  \\[3mm]
\displaystyle \frac{du_i}{dt} & = & \kappa_{u_i}\left(1 - s^-(x_{i})s^-(x_{i+1})\right) - \gamma_{u_i}u_i\,.\\[3mm]
\end{array}
\end{equation}

%

Let us also assume that the units are identical, so that in the $4n$ model~\eqref{eq:4n}, $\kappa_{x_i}=\kappa_1$, $\kappa_{y_i}=\kappa_2$, $\kappa_{z_i}=\kappa_3$, and $\kappa_{u_i}=\kappa_4$ for each unit $i \in \{1,\ldots , n\}$, and in the $2n$ model, $\kappa_{x_i}=\kappa_1$, $\kappa_{u_i}=\kappa_2$. The same is true for the thresholds.
In what follows, we will also take all decay rates to be the same, so that $\gamma_{x_i}=\gamma_{y_i}=\gamma_{z_i}=\gamma_{u_i}=\gamma$. 
In this case, a rescaling of the time coordinate allows us to take $\gamma=1$. In the $2n$ model, however, we are implicitly assuming that the inverters ($y_i$ and $z_i$) are infinitely fast.
%

Again we can translate thresholds to the origin in the dimension $2n$ model, so that we get the simpler form, written with a $v$ notation as for the $4n$ case:
\begin{equation}
\begin{array}{lcl}
\displaystyle \frac{dv_{2i-1}}{dt} & = &  \kappa_{1} (s^+(v_{2i-3})s^-(v_{2i})+s^-(v_{2i-3})s^+(v_{2i}) -  \theta _1 - v_{2i-1} \\[3mm]
\displaystyle \frac{dv_{2i}}{dt}   & = &  \kappa_{2} (1-s^-(v_{2i-1}))s^-(v_{2i+1})-  \theta _2 -  v_{2i}\,,
\end{array}
\label{eq:2n2v2}
\end{equation}
where now $(v_{2i-1},v_{2i})=(x_i-\theta_1,u_i-\theta_2) $.

Before attempting analysis of these models, we first note that systems of the form \eqref{eq:4n}-\eqref{eq:2n2v2} all belong to the class of piecewise-linear (PL) differential equations known as Glass networks \cite{glass1975combinatorial,glass2018hybrid,glass1973logical}, so we provide a brief introduction to the analysis of dynamics in such systems in the next sections, emphasizing periodic orbits and their bifurcations.  

\section{Background on Glass networks}\label{sec:glassnets}

There is now a fairly extensive literature on Glass networks, spanning several decades, but most of the tools summarized in the remainder of this section and the next are found in~\cite{edwards2000analysis} and \cite{killough2005bifurcations}. For this discussion, we use a notation general enough to encompass all models investigated in this paper; compared to more general Glass networks, it will be assumed hereafter that all decay rates are equal, and that thresholds have been set to zero as in~\eqref{eq:2n2v2}. Let us then write the general form: 
\begin{equation}\label{eq:Glass}
\dot v = S(v) - \gamma v,
\end{equation}
where $v\in\mathbb{R}^N$, $\gamma>0$, and $S$ is expressed in terms of Heaviside functions as above, with $0$ thresholds, in such a way that for any $v$ such that $v_i\ne 0$, one has $S_i(v)\in\{\kappa_i^-,\kappa_i^+\}$, where $-\gamma\theta_i=\kappa_i^-<0<\kappa_i^+=\kappa_i-\gamma\theta_i$. 

The system is dissipative, due to the $-\gamma v$ term, and from this one can show that all trajectories will enter the rectangular domain $R=\prod_i[\kappa_i^-,\kappa_i^+]$ in finite time, so the latter can be considered a state space for this model. Then, due to the occurrence of Heaviside functions there is a natural partition of the state space $R$ into $N$-dimensional cuboids $\prod_i I_i$ where $I_i$ are intervals of the form $[-\kappa_i^-,0)$ or $(0,\kappa_i^+]$. Such domains will be termed \textit{boxes} and we denote them using a Boolean vector $b=(b_i)$, where $b_i$ is zero (resp. one) for $I_i$ being the lower (resp. higher) interval. We abuse notation and use $b$ to refer either to a Boolean vector, or to the sub-domain in $R$ as required by the context. 

In each ``regular domain'' $b\in\{0,1\}^N$, the term $S(v)$ is constant and therefore can be denoted as $S(b)$ without ambiguity. In $b$, the dynamics is purely linear and converges to a steady state $\phi(b)=S(b)/\gamma$, traditionally termed the ``focal point'' for $b$. If this focal point is in the domain $b$ it is an asymptotically stable steady state, and otherwise all trajectories starting in $b$ exit this domain in finite time. In this case, and as detailed in \cite{edwards2000analysis}, one can define a mapping from points in the closure of $b$ to its boundary, which associates to each point $v$ the corresponding exit point, {\em i.e.}, 
the first $v(t)$ for $t>0$ where the solution curve starting at $v(0)$ hits the boundary of $b$. 
In fact, choosing a time scale so that $\gamma=1$, the solution of the system within the box is
\begin{equation}
    v(t)=\phi(b)+\left(v(0)-\phi(b)\right)e^{-t}\,.
\end{equation}
If the trajectory hits the boundary where $v_j(t^*)=0$, then $0=\phi_j(b)+\left(v_j(0)-\phi_j(b)\right)e^{-t^*}$, so $e^{-t^*}=\frac{\phi_j(b)}{\left(\phi_j(b)-v_j(0)\right)}$, and the exit point of the box is 
\begin{equation}\label{eq:GlassOneStep}
\begin{aligned}
    v(t^*)&=\phi(b)+\left(v(0)-\phi(b)\right)\frac{\phi_j(b)}{\left(\phi_j(b)-v_j(0)\right)}\\ &=
    \frac{v(0)\phi_j(b) - \phi(b)v_j(0)}{\left(\phi_j(b)-v_j(0)\right)}=\frac{v(0)-\frac{\phi(b)}{\phi_j(b)}v_j(0)}{1-\frac{v_j(0)}{\phi_j(b)}}\,.
    \end{aligned}
\end{equation}
Conveniently, this takes the form of a fractional linear mapping, and can be formulated in terms of a matrix $B$ and vector $\psi$ defined in terms of focal point coordinates:
\begin{equation}
\label{eq:exitmap}
Mv = \frac{B v}{1+\psi^\top v},
\end{equation}
where the superscript $^\top$ denotes transpose. This form is preserved under composition and therefore also describes the mapping from $v$ to its image after an arbitrary number of threshold intersections, along the solution curve starting from $v$. A more specific notation will be required in subsequent sections. Assume a solution curve with initial condition $v^{(0)}\in b^{(0)}$, and which successively crosses $m$ boxes $b^{(k)}$, $1\leq k\leq m$, at locations $v^{(k)}$ on the intersection of the boundaries of $b^{(k-1)}$ and $b^{(k)}$. It is generically expected that at each exit point a single threshold is crossed, so that $b^{(k)}$ and $b^{(k+1)}$ differ at a single ``switching coordinate'' $j=j(k)$ ($k$ will be omitted when clear in context), hence $v^{(k+1)}_{j(k)}=0$. The successive focal points are denoted $\phi^{(k)}=\phi(b^{(k)})$. Then, the mapping from the boundary of $b^{(k)}$ to that of $b^{(k+1)}$ takes the form \eqref{eq:exitmap} with, in the numerator, the matrix
\begin{equation}\label{eq:Bk}
B^{(k)} = I - \frac{1}{\phi^{(k)}_{j(k)}} \phi^{(k)}\mathbf{e}_{j(k)}^\top = 
\scalebox{.8}{$\left(
\begin{array}{ccccccc}
1 &        & & -\phi^{(k)}_1/\phi^{(k)}_j & & & \\
  & \ddots & & \vdots && \\
  &        & 1 & -\phi^{(k)}_{j-1}/\phi^{(k)}_j & &  &\\[2mm]
 &&&0&&& \\[1mm]
 && & -\phi^{(k)}_{j+1}/\phi^{(k)}_j &1  & \\
 &&& \vdots && \ddots & \\
 && & -\phi^{(k)}_N/\phi^{(k)}_j & & & 1\\
\end{array}
\right)$}
\end{equation}
(See~\cite{edwards2000analysis} for the derivation). 
The $j^{th}$ row is zero and all columns but the $j^{th}$ are standard basis (column) vectors, above denoted $\mathbf{e}_{j}$ for the $j^{th}$. The vector in the denominator of \eqref{eq:exitmap} is, for the $k^{th}$ step:
\begin{equation}\label{eq:psik}
\psi^{(k)} = \frac{-1}{\phi^{(k)}_{j(k)}}\mathbf{e}_{j(k)}.
\end{equation}
Direct calculation shows that the mapping from $b^{(0)}$ to the boundary of $b^{(m)}$ is of the same form \eqref{eq:exitmap}, with matrix and vector:
\begin{equation} \label{eq:B0m}
\begin{aligned}
B^{(m,0)} =& B^{(m-1)}B^{(m-2)}\cdots B^{(0)}\quad\text{and}\; \\\psi^{(m,0)}= &
\psi^{(0)}+\sum_{k=1}^{m-1}B^{(k,0)\top}\psi^{(k)}
\end{aligned}
\end{equation}
respectively, where the full denominator can also be written as $1+\psi^{(m,0)\top}v^{(0)} = \prod_{k=0}^{m-1}\left(1+\psi^{(k)\top}v^{(k)}\right)$, which is $\exp({t^{(m)}})$, the exponential of the time taken in following these $m$ steps (see~\cite{edwards2000analysis}) for details). 

Importantly, the domain of a mapping such as the above has to be specified. Indeed if along the sequence of boxes one box has multiple possible successors (which occurs exactly when the box containing $\phi(b)$ differs from $b$ at multiple digits), then one mapping can be defined for each possible switching coordinate. Again referring to \cite{edwards2000analysis} for details, consider that box $b^{(k-1)}$, with switching variable $j$ to box $b^{(k)}$, also has an alternative switching variable $i$, leading to another successor box than $b^{(k)}$. The set of initial conditions $v^{(k)}$ in $b^{(k)}$ for which $v_j^{(k)}=0$ before $v_i^{(k)}=0$, {\em i.e.}, for which the solution exits $b^{(k-1)}$ towards $b^{(k)}$ rather than the alternative, can be written as 
\[
\frac{-\mathbf{e}_i^T}{\phi_i^{(k)}}B^{(k)} v^{(k)}>0, 
\]
where $B^{(k)}$ is exactly as in \eqref{eq:Bk}, {\em i.e.}, with switching coordinate $j$. Because the denominators $1+\psi^{(k)\top}v^{(k)}$ can be shown to be positive (they are, in fact, the exponential of the time spent going from $v^{(k)}$ to $v^{(k+1)}$), the above is equivalent to an inequality in $v^{(0)}$, namely:
\begin{equation} \label{eq:switchineq}
\frac{-\mathbf{e}_i^\top}{\phi_i^{(k)}}B^{(k)}B^{(k-1)}\cdots B^{(0)} v^{(0)}>0.
\end{equation}
Note that the $\mathbf{e}_i^\top$ factor means that the above is a single (scalar) inequality, of the form $R_iv^{(0)}>0$ for a row vector $R_i$. Given the generic sequence we are discussing now, for any box $b^{(k)}$ having alternative switching variables besides $j(k)$, there will be one row $R_i$ for each alternative $i$. Combining all these rows (in an arbitrary order), for all alternative switching variables and all boxes along the sequence, produces a matrix we shall denote by $R=R(b^{(0)},b^{(1)},\dots,b^{(m)})$. Then, the inequality
\begin{equation}\label{eq:Rx>0}
Rv>0,
\end{equation}
where the inequality is interpreted to apply to all components, defines exactly the set of initial conditions $v^{(0)}$ whose trajectory starts with the required sequence of boxes. The inequality above defines the interior of a proper (polyhedral) cone, hereafter denoted by $C$. In practice, many of the rows in matrix $R$ are redundant (linearly dependent) and its computation can be simplified accordingly, but there is no danger in keeping all rows. 

A key result, for the purpose of this paper, is that the formulation above allows one to characterize the existence and stability of periodic orbits, in terms of elementary linear algebra. At the core of this approach is the interpretation of the mapping \eqref{eq:exitmap}, considered for a cyclic sequence of boxes, as a Poincaré map. The associated Poincaré section is the intersection of the cone \eqref{eq:Rx>0} for this sequence of boxes and the interface (or ``wall'') between the first two boxes. This allows in particular a description of bifurcations through which limit cycles appear or disappear, coalesce, or change stability. A classification of these bifurcations was constructed in~\cite{killough2005bifurcations}, which in the present context will provide a basis to describe how chaos may arise in models such as \eqref{eq:4n}-\eqref{eq:2n2v2}.

This classification is built upon the characterization of limit cycles proved in \cite{edwards2000analysis}. Consider a model of the form \eqref{eq:Glass} and assume that some trajectories follow a particular cyclic sequence of boxes $b^{(0)},\dots,b^{(m-1)},b^{(m)}=b^{(0)}$. The return map of that cycle is of the form \eqref{eq:exitmap}, with $B^{(m,0)}$ and $\psi^{(m,0)}$ calculated as in \eqref{eq:B0m}. Then, there is a periodic orbit through this cycle of boxes if and only if the return map has both of the following ``limit cycle'' properties:
\begin{enumerate}
\item[\textsf{LC1}.] A real eigenvector $w_i$ of $B^{(m,0)}$ lies in the returning cone for the cycle ({\em i.e.}, $Rw_i>0$).
\item[\textsf{LC2}.] The corresponding eigenvalue is real and satisfies $\lambda_i > 1$.
\end{enumerate}
The periodic orbit is stable if and only if the return map satisfies the additional condition:
\begin{enumerate}
\item[\textsf{LC3}.] $\lambda_i \geq |\lambda_j|\quad\text{ for all }j \ne i$.
\end{enumerate} 
The periodic orbit is asymptotically stable if strict inequality holds in the third condition. The fixed point of the return map corresponding to the periodic orbit is in fact a particular multiple of the eigenvector $w_i$, given by
\begin{equation}\label{eq:ystar}
v^* = \frac{(\lambda_i-1)w_i}{\psi^\top w_i}.
\end{equation}

Note that one row of $B^{(m,0)}$ is always $0$, say the $k^{th}$, and for a cycle map the corresponding component of the vector $v^{(0)}$ is also $0$, {\em i.e.}, the $k^{th}$ component is zero initially on the starting wall, and mapped back to zero on return to the starting wall. Therefore, the map can be reduced to dimension $N-1$ by removing the $k^{th}$ row and column of $B^{(m,0)}$, and the $k^{th}$ component of $\psi^{(m,0)}$ and $v$. This reduces the map for a cycle to $M:\R^{N-1}\to\R^{N-1}$ with matrix $B\in\R^{(N-1)\times(N-1)}$ and vector $\psi\in\R^{N-1}$, but in this paper we keep the extra variable (except where specifically noted), so that $M:\R^{N}\to\R^{N}$, with $B\in\R^{N\times N}$ and $\psi\in\R^{N}$:
\begin{equation}
    M(v) = \frac{B v}{1+\psi^\top v}
    \label{eq:cyclemap}
\end{equation}
where $v$ is a vector in $\R^{N}$, one of whose components will be $0$.
This implies that $B$ will always have a $0$ eigenvalue and corresponding eigenvector. 

Note that the period of a periodic orbit through a fixed point of this map is $T=\log(\lambda)$ where $\lambda=1+\psi^{\top}v^*$ is the eigenvalue of $B$ corresponding to eigenvector $v^*$, since $M(v^*)=v^*\implies Bv^*=(1+\psi^{\top}v^*)v^*$.

\section{Background on bifurcations of cycles}\label{sec:bifurcations}

\begin{widetext}
The possible bifurcations of limit cycles, as classified by Killough and Edwards~\cite{killough2005bifurcations}, are listed below. We assume that we have a limit cycle on one side of the bifurcation. The symbol $\mathcal{S}$ denotes the list of variables that switch along the cycle.
\\

\noindent
\textbf{DS: Double-switching bifurcation}: $v^* \in\partial C$, and $v^*_j\ne 0$ for some $v_j\in \mathcal{S}$.\\

\noindent
\textbf{CC: Cycle-collapse bifurcation}: $\lambda_i=1$ (in which case $v_j^*=0$ for all $j\in\mathcal{S}$, by~\eqref{eq:ystar}). 

\noindent
\textbf{CD: Cycle-destabilizing bifurcation}: $\lambda_i=|\lambda_j|>1$ for some $j\ne i$, and $\lambda_i>|\lambda_k|$ for all $k\ne i,j$.

\hfill\\ \begin{minipage}{\dimexpr\textwidth-1cm} \begin{itemize}
\item[(a)] At the bifurcation  $\lambda_i=\lambda_j$, and after the bifurcation $\lambda_i<\lambda_j$ (both real).
\item[(b)] At the bifurcation  $\lambda_i=\lambda_j$, and after the bifurcation $\lambda_i=\bar \lambda_j$ (both complex).
\item[(c)] At the bifurcation  $\lambda_i= -\lambda_j$, and after the bifurcation $\lambda_i<-\lambda_j$.
\item[(d)] At the bifurcation  $\lambda_i=|\lambda_j|=|\lambda_k|$, $\lambda_j=\bar \lambda_k$, and after the bifurcation $\lambda_i<|\lambda_j|=|\lambda_k|$.
\end{itemize}\end{minipage}\ \\
\\

\noindent
\textbf{S: Structural bifurcation}: $\phi_i = 0$ where $\phi_i$ is the $i^{th}$ component of the focal point associated with some box on the cycle.\\
The above definitions specify the condition that holds exactly at the bifurcation point. Of course, in order for the bifurcation to occur as a parameter is changed, one must pass through such a bifurcation point transversally.

\end{widetext}

\begin{figure}
\begin{center}
\includegraphics[width=.6\columnwidth]{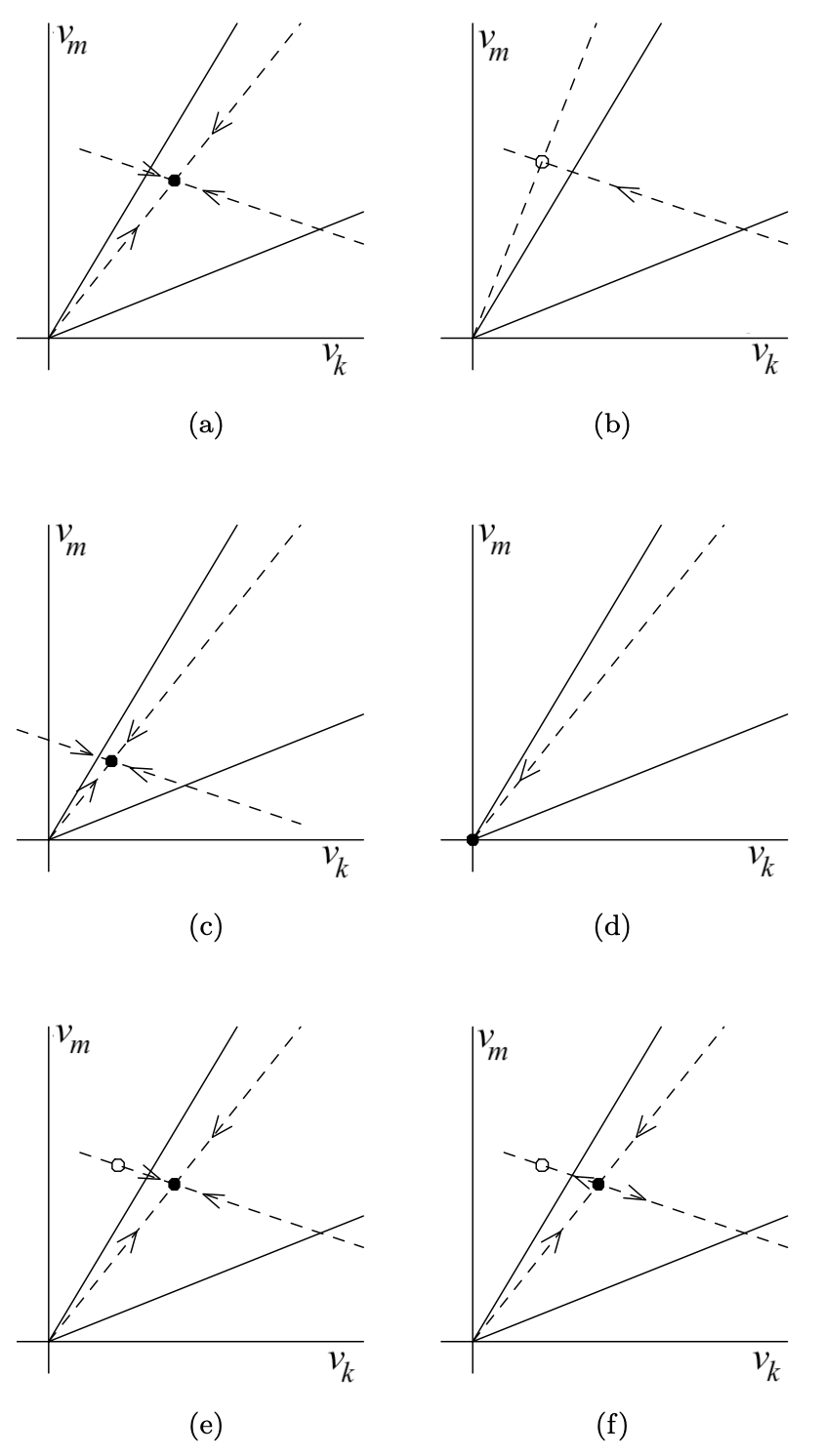}
\end{center}
\caption{Bifurcation types depicted on a $2$-dimensional wall in a $3$-dimensional network (or, equivalently, on a projection of an $(N-1)$-dimensional wall in an $N$-dimensional network). Poincar\'e sections in the positive quadrant in the plane of these two variables are shown before and after a double-switching bifurcation in (a) and (b), a cycle-collapse bifurcation in (c) and (d) and a cycle-destabilizing bifurcation of type `a' in (e) and (f). The solid lines indicate the boundaries of the returning cone for a cycle. Fixed points of the Poincar\'e map are indicated by circles, filled if inside the returning cone, open if outside. Dotted lines indicate eigenvectors of the linearized map at the fixed point and arrows show stability.
}
\label{fig:bifurcation_types}
\end{figure}

Figure~\ref{fig:bifurcation_types} (adapted from Killough and Edwards~\cite{killough2005bifurcations}) shows sketches of $DS$, $CC$, and $CD(a)$ bifurcations in a $2$-dimensional wall of a system with $N=3$, or equivalently, a projection of an $(N-1)$-dimensional wall in an $N$-dimensional system.

Killough and Edwards~\cite{killough2005bifurcations} give results on how to determine, for each type of bifurcation, what cycles, stable or unstable, exist on either side of the bifurcation. We summarize these results here, as we will use them later. The results on {\bf DS} bifurcations originate from an analysis of non-smooth bifurcations by Feigin~\cite{Feigin1995} and di Bernardo, {\em et al.}~\cite{diBernardo1999}, but are adapted by Killough and Edwards for Glass networks. For simplicity, we will assume that the conditions for only one type of bifurcation occur at the bifurcation point. It is possible for more than one of these conditions to occur simultaneously, and more complicated bifurcations scenarios may then arise, but we will not need to consider such situations here.

A {\bf CC} bifurcation occurs for a cycle when at some parameter value $\lambda_i=1$, if on one side of the bifurcation point $\lambda_i>1$ and on the other side $\lambda_i<1$, while the corresponding eigenvector, $v_i$, lies in the cycle's returning cone on both sides. If $\lambda_i$ is dominant, the cycle is stable on one side, and collapses to a fixed point on the other, which is at least semi-stable. Otherwise, both the cycle and the resulting fixed point are unstable. We can represent these bifurcations of cycles as $A\to \emptyset$ or $a\to\emptyset$, where by convention, the capital $A$ implies that cycle $A$ is stable, while the lower-case $a$ implies that it is unstable, and the $\emptyset$ means that there is no cycle after the bifurcation.

A bifurcation of type {\bf CD(a)} occurs when real eigenvalues swap dominance. The cycle is stable before the bifurcation and unstable after. If the fixed point, ${\bf v}_j^*$, associated with eigenvalue $\lambda_j$ is in the returning cone for the cycle, then a previously unstable cycle becomes stable after the bifurcation. In that case, we have $A,b\to a,B$. If ${\bf v}_j^*$ is not in the returning cone, we have $A\to a$.

{\bf CD(b)} bifurcations are $A,b\to \emptyset$.

{\bf CD(c)} bifurcations are $A\to a$. No period-doubling can occur.

{\bf CD(d)} bifurcations are also $A\to a$.

{\bf S} bifurcations are $A\to \emptyset$ or $a\to \emptyset$. A periodic orbit is lost and a fixed point is gained.
\\

\begin{figure}%
\begin{center}
\includegraphics[width=.8\columnwidth]{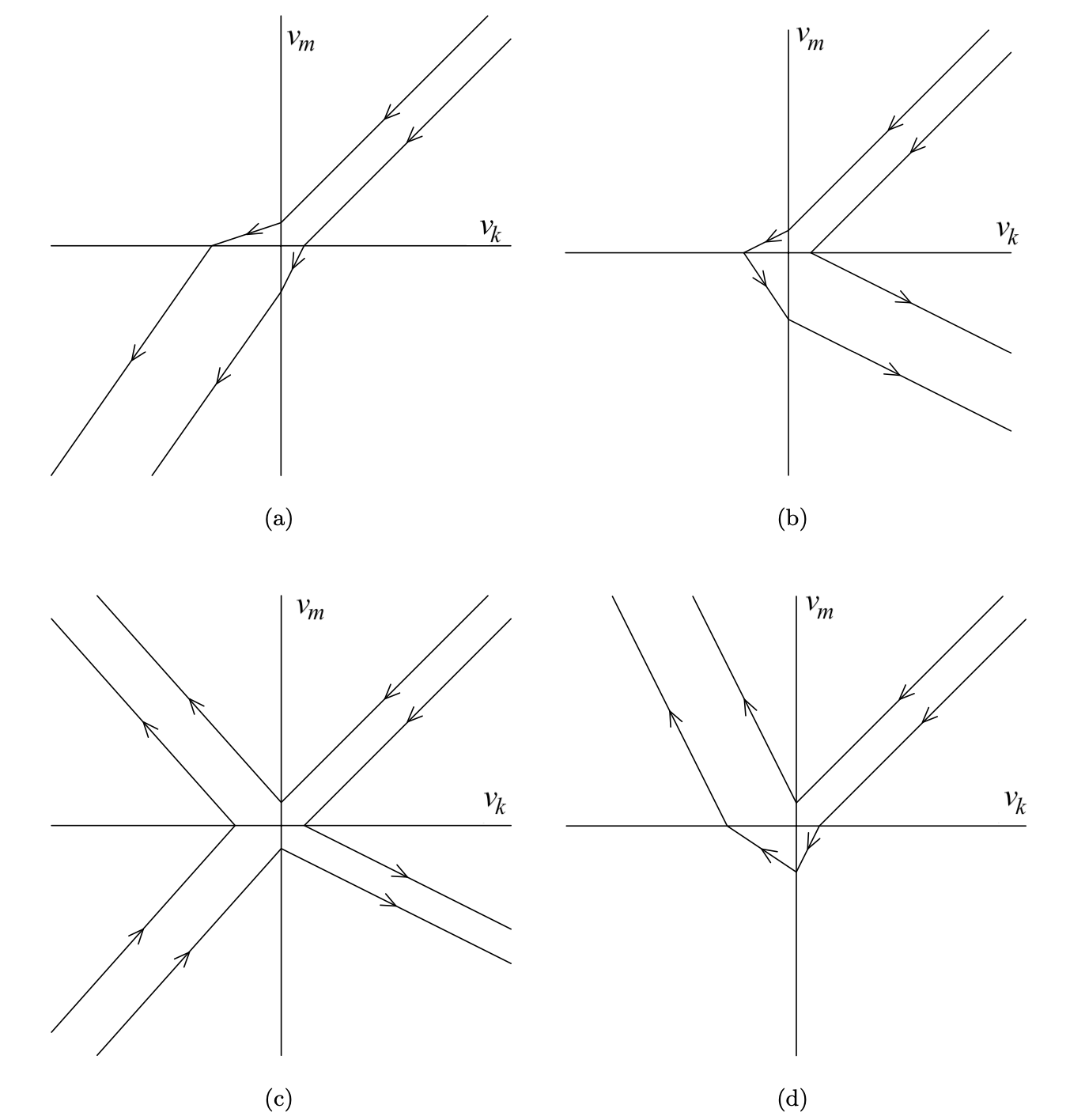}
\end{center}
\caption{Flow diagrams in the plane of the two variables, $v_k$ and $v_m$, involved in a double-switching bifurcation. The focal point coordinates for the upper right quadrant are in the lower left quadrant (other cases are equivalent by symmetry). There are four cases for the focal point coordinates of the lower left quadrant. (a) $\phi_k < 0,\; \phi_m < 0$; unambiguous (b) $\phi_k > 0,\; \phi_m < 0$; unambiguous (c) $\phi_k > 0,\; \phi_m > 0$; ambiguous (d) $\phi_k< 0,\; \phi_m > 0$; unambiguous.
}
\label{fig:doubleswitching}
\end{figure}

{\bf DS} bifurcations need to be further classified. We assume existence of a stable cycle before the bifurcation. If trajectories are continuously deformed across the bifurcation, then after the bifurcation there is a new cycle with a different sequence of boxes, where the two variables that switch simultaneously somewhere around the cycle at the bifurcation value of the parameter, either switch in the opposite order, or two new switches of one of the two variables are either gained or lost. This case is called unambiguous double-switching by Killough and Edwards~\cite{killough2005bifurcations} and is clear from Figure~\ref{fig:doubleswitching}, adapted from that paper. In one possible arrangement of focal points, however, the trajectory changes discontinuously at the bifurcation point, and there may be no cycle after the bifurcation. This is called ambiguous double-switching.

{\bf DS(a)}: Ambiguous {\bf DS} bifurcations are $A\to\emptyset$.

For unambiguous {\bf DS} bifurcations, the existence and stability of cycles on either side is determined by properties of the eigenvalues of the matrix $A$ for the cycle before the bifurcation, and of matrix $B$ for the cycle after the bifurcation. Let $\alpha_j,\, j=1,\ldots, N$ be the eigenvalues of the matrix $A$, and $\beta_j,\, j=1,\ldots, N$ be the eigenvalues of the matrix $B$, in both cases evaluated at the bifurcation point. We assume that we have a periodic orbit before the bifurcation through the fixed point ${\bf v}_i^*$ associated with eigenvalue $\alpha_i>1$, and a periodic orbit associated with the new cycle after the bifurcation through the fixed point ${\bf w}_i^*$ associated with eigenvalue $\beta_i>1$. Note that at the bifurcation point, ${\bf v}_i^* = {\bf w}_i^*$.

Define
\begin{itemize}
\item $\sigma_\alpha^+=$ the number of real eigenvalues of $A$ greater than $\alpha_i$;
\item $\sigma_\beta^+=$ the number of real eigenvalues of $B$ greater than $\beta_i$;
\item $\sigma_\alpha^-=$ the number of real eigenvalues of $A$ less than $-\alpha_i$;
\item $\sigma_\beta^-=$ the number of real eigenvalues of $B$ less than $-\beta_i$.
\end{itemize}
Define $\sigma_{\alpha\beta}^+$ and $\sigma_{\alpha\alpha}^*$ analogously for the cycle compositions $AB$ and $AA$, respectively.
Note that under our assumption that cycle $A$ has a stable periodic orbit before (and up to) the bifurcation, we must have $\sigma_\alpha^+=\sigma_\alpha^- = 0$, but these results are more general.

Now, whe have the following result~\cite{killough2005bifurcations}:

\begin{prop} \label{prop:DSbif}
For an unambiguous {\bf DS} bifurcation,

\begin{itemize}
\item {\bf DS(b)}: If $\sigma_\alpha^- + \sigma_\beta^-$ is even, and $\sigma_\alpha^+ + \sigma_\beta^+$ is even, we have $A\to B$;
\item {\bf DS(c)}: If $\sigma_\alpha^- + \sigma_\beta^-$ is even, and $\sigma_\alpha^+ + \sigma_\beta^+$ is odd, we have $A,b\to \emptyset$;
\item {\bf DS(d)}: If $\sigma_\alpha^- + \sigma_\beta^-$ is odd, $\sigma_\alpha^+ =\sigma_\alpha^-=0$, $\sigma_\beta^+$ is even, and $\sigma_{\alpha\beta}^+ + \sigma_{\alpha\alpha}^+ =0$, we have $A\to b,AB$;
\item {\bf DS(e)}: If $\sigma_\alpha^- + \sigma_\beta^-$ is odd, $\sigma_\alpha^+ =\sigma_\alpha^-=0$, $\sigma_\beta^+$ is even, and $\sigma_{\alpha\beta}^+ + \sigma_{\alpha\alpha}^+ \ne 0$ but is even, we have $A\to b,ab$;
\item {\bf DS(f)}: If $\sigma_\alpha^- + \sigma_\beta^-$ is odd, $\sigma_\alpha^+ =\sigma_\alpha^-=0$, $\sigma_\beta^+$ is even, and $\sigma_{\alpha\beta}^+ + \sigma_{\alpha\alpha}^+$ is odd, we have $A,ab\to b$;
\item Other cases not needed here (see~\cite{diBernardo1999,Feigin1995}).
\end{itemize}
\end{prop}

These behaviours can be tabulated as in Table~\ref{tab:biftypes} (assuming a stable periodic orbit before the bifurcation).

\begin{table}
\begin{center}
\begin{tabular}{|l|l|}
\hline
Type & Behaviour \\ \hline
{\bf CC}  & $A\to\emptyset$ \\
{\bf CD(a)} & $A,b\to a,B$ or $A\to a$ \\
{\bf CD(b)} & $A,b\to \emptyset$ \\
{\bf CD(c)} & $A\to a$ \\
{\bf CD(d)} & $A\to a$ \\
{\bf S} & $A\to \emptyset$ \\
{\bf DS(a)} & $A\to \emptyset$ \\
{\bf DS(b)} & $A\to B$ \\
{\bf DS(c)} & $A,b\to \emptyset$ \\
{\bf DS(d)} & $A\to b,AB$ \\
{\bf DS(e)} & $A\to b,ab$ \\
{\bf DS(f)} & $A,ab\to b$ \\
\hline
\end{tabular}
\caption{ Types of bifurcations of periodic orbits.}\label{tab:biftypes}
\end{center}
\end{table}

\vspace{2mm}
In Section~\ref{sec:4n}, we aim to illustrate how these tools can be used to describe in great detail how chaos arises in a high-dimensional model of the form \eqref{eq:2n2v2}, via a series of bifurcations of some of the types above.

\section{Analysis of a $10$-dimensional model of the ring circuit} \label{sec:10d}

In order to illustrate the analysis of limit cycles described in Section~\ref{sec:glassnets}, we start with the simplified model of the ring circuit, in only $10$ dimensions, where it is not overly cumbersome to display the matrices and vectors involved in the calculations.
We thus consider in this section the $2n$ model~\eqref{eq:2n2v2} with $5$ units $(n=5)$, and $\theta_1=\theta_2=\theta$ and $0<\theta <\min\{\frac{\kappa_1}{\gamma}, \frac{\kappa_2}{\gamma}\}$, to ensure that switching is possible. Without losing further generality, we will always take $\theta =\frac{1}{2}$ to set a size scale, $\gamma=1$ to set a time scale, and allow $\kappa_1$ and $\kappa_2$ to vary in the appropriate range: $\min(\kappa_1,\kappa_2)>\frac 12$.

Numerically, we observe the following cycle of states occurring for a range of parameter values and initial conditions chosen arbitrarily, but satisfying the above constraints:
\begin{equation}
\begin{array}{lllll}
\multicolumn{5}{c}{\text{state}} \\ \hline 
      01 &11 &11 &01 &11  \\
      01 &11 &01 &01 &11  \\
      01 &11 &01 &11 &11  \\
      01 &11 &01 &11 &01  \\
      11 &11 &01 &11 &01  \\
      11 &01 &01 &11 &01  \\
      11 &01 &11 &11 &01  \\
      11 &01 &11 &01 &01  \\
      11 &01 &11 &01 &11  \\
      01 &01 &11 &01 &11 
\end{array}\hspace{1cm}
\begin{array}{lllll}
\multicolumn{5}{c}{\text{focal point box}} \\ \hline 
      01 &11 &{\bf 0}1 &01 &11  \\
      01 &11 &0{\bf 0} &{\bf 1}1 &11  \\
      01 &11 &01 &11 &{\bf 0}1  \\
      {\bf 1}1 &11 &01 &11 &0{\bf 0}  \\
      11 &{\bf 0}1 &01 &11 &01  \\
      11 &0{\bf 0} &{\bf 1}1 &11 &01  \\
      11 &01 &11 &{\bf 0}1 &01  \\
      11 &01 &11 &0{\bf 0} &{\bf 1}1  \\
      {\bf 0}1 &01 &11 &01 &11  \\
      0{\bf 0} &{\bf 1}1 &11 &01 &11 
\end{array}
\label{eq:Cycle}
\end{equation}

The focal points for each of the states visited are listed at right in terms of the box in which they sit. The potential switching coordinates are shown in bold font. The actual focal point coordinates are $-\theta=-\frac 12$ if the box coordinate is $0$, and $\frac{\kappa_i}{\gamma}-\theta=\kappa_i-\frac 12$ if the box coordinate is $1$, where $i=1$ for odd coordinates, and $i=2$ for even coordinates.
The sequence of variables (indices) that undergo transitions at each step (step 1 to step 10) is: 5, 7, 9, 1, 3, 5, 7, 9, 1, 3. Note that at each of the even steps  there is an alternate exit variable that is not followed on the cycle, namely variables 6, 10, 4, 8 and 2 at steps 2, 4, 6, 8, and 10, respectively,

The maps~\eqref{eq:exitmap}, \eqref{eq:Bk}, \eqref{eq:psik} for each step of the cycle can be calculated explicitly. For example, after the first step, during which $v_5$ switches, we have $v_1^{(1)}=\frac{v_1-v_5}{1+2v_5}$, where the right-hand side is evaluated at step $0$, and $v_2^{(1)}=\frac{v_2-v_5+2\kappa_2 v_5}{1+2v_5}$, etc. These come from Equation~\eqref{eq:GlassOneStep}. Of course, $v_5^{(1)}=0$. Writing the entire map for the first step in vector-matrix form, 
\begin{equation} v^{(1)}=\frac{B^{(0)}v^{(0)}}{1+\psi^{(0)\top}v^{(0)}} \end{equation}
where $v=(x_1-\theta \quad u_1-\theta \quad x_2-\theta \quad u_2-\theta \quad \ldots \quad x_5-\theta \quad u_5-\theta)^{\top}$, we get
\begin{equation}
B^{(1)}=
\begin{pmatrix} 
 1& 0& 0& 0&              -1 & 0& 0& 0& 0& 0\\
 0& 1& 0& 0& 2\kappa_2-1& 0& 0& 0& 0& 0\\
 0& 0& 1& 0& 2\kappa_1-1& 0& 0& 0& 0& 0\\
 0& 0& 0& 1& 2\kappa_2-1& 0& 0& 0& 0& 0\\
 0& 0& 0& 0&               0& 0& 0& 0& 0& 0\\
 0& 0& 0& 0& 2\kappa_2-1& 1& 0& 0& 0& 0\\
 0& 0& 0& 0&              -1& 0& 1& 0& 0& 0\\
 0& 0& 0& 0& 2\kappa_2-1& 0& 0& 1& 0& 0\\
 0& 0& 0& 0& 2\kappa_1-1& 0& 0& 0& 1& 0\\
 0& 0& 0& 0& 2\kappa_2-1& 0& 0& 0& 0& 1\\
\end{pmatrix},\quad \psi^{(1)}=
\begin{pmatrix} 
0\\ 0\\ 0\\ 0\\ 2\\ 0\\ 0\\ 0\\ 0\\ 0
\end{pmatrix} ,
\label{eq:firststep}
\end{equation}
and similarly for the other 9 steps, which can then be composed to get a return map for the cycle.

The presence of $5$ alternate exit variables around the cycle implies that the returning cone for the cycle is not the entire starting wall. 
The returning cone, using~\eqref{eq:switchineq} or \eqref{eq:Rx>0}, is given by a $5\times 10$ matrix $R$, which depends on values of $\kappa_1$ and $\kappa_2$. The cycle maps and the matrix $R$ that defines the returning cone can be computed symbolically in terms of parameters. However, the theorems on existence and stability of periodic orbits, as well as bifurctions of these, all require calculation of eigenvalues and eigenvectors, which cannot be done symbolically.

To illustrate with particular parameter values, let 
\begin{equation}\label{eq:kappas}
(\kappa_1, \kappa_2)=(1.53,2).
\end{equation}
Then the map, $M$, on $\R^{10}$ is given by

\begin{widetext}
\begin{equation*}
B=
\begin{pmatrix*}[r]
  13.691830  & 0.0  & -9.5441731  & 0.0  & -177.13291  & 0.0  &  58.957699  & 0.0  & -41.175914  & 0.0 \\
 -75.37999  & 1.0  &  52.438817  & 0.0  &  979.53145  & 0.0  & -325.65676  & 0.0  &  226.92754  & 0.0 \\
       0.0  & 0.0  &        0.0  & 0.0  &        0.0  & 0.0  &        0.0  & 0.0  &        0.0  & 0.0 \\
-127.26297  & 0.0  &  88.673761  & 1.0  &  1649.8864  & 0.0  & -548.37695  & 0.0  &  382.80511  & 0.0 \\
-84.822382  & 0.0  &  59.162649  & 0.0  &   1096.9140  & 0.0  & -364.89379  & 0.0  &  255.01105  & 0.0 \\
-130.14731  & 0.0  &  90.615509  & 0.0  &  1686.1213  & 1.0  &  -559.54580  & 0.0  &   391.63120  & 0.0 \\
 36.630477  & 0.0  & -25.659733  & 0.0  & -473.31987  & 0.0  &  157.47191  & 0.0  & -109.88274  & 0.0 \\
-117.03671  & 0.0  &  81.789418  & 0.0  &  1523.3594  & 0.0  & -506.72023  & 1.0  &  353.45451  & 0.0 \\
-68.706823  & 0.0  &  48.313613  & 0.0  &  890.21656  & 0.0  & -296.18697  & 0.0  &  206.69744  & 0.0 \\
-128.20556  & 0.0  &  90.615509  & 0.0  &  1679.2370  & 0.0  & -558.60321  & 0.0  &  389.68945  & 1.0 \\
\end{pmatrix*},
\end{equation*}
\end{widetext}

\begin{equation*}
\psi=
\begin{pmatrix}
 -86.764873 \\ 0\\ 60.410339\\ 0\\ 1125.3754\\ 0\\ -374.32503\\ 0\\ 261.08747\\ 0
\end{pmatrix} .
\end{equation*}
The eigenvalues of the matrix $B$ are the components of the vector
\begin{equation*}
\lambda=
\begin{pmatrix} 
             1474.96  \\       
                 1.0  \\
                 1.0  \\
                 1.0  \\
                 1.0  \\
                 1.0  \\
 - 0.115664 - 0.023963i  \\
 - 0.115664 + 0.023963i  \\
              0.048593  \\
                0.0 \\
\end{pmatrix}.
\end{equation*}
Note that the dominant eigenvalue is $\lambda_{\max}=1474.96$, where the corresponding eigenvector is given by  
\begin{equation*}
w_{\max}=
\begin{pmatrix} 
 -0.047475\\
  0.262541\\
  0 \\
  0.442304\\
  0.293978\\
  0.452013\\
 -0.126830\\
  0.408405\\
  0.238541\\
  0.450183
\end{pmatrix}.
\end{equation*}
The fixed point for this cycle is given by
\begin{equation} \label{eq:fixedpoint10}
v^*= \frac{(\lambda_{\max}-1)}{\psi^{(10,0)\top} w_{\max}}w_{\max} = 
\begin{pmatrix} 
v_1\\
v_2\\
v_3\\
v_4\\
v_5\\
v_6\\
v_7\\
v_8\\
v_9\\
v_{10}
\end{pmatrix} = 
\begin{pmatrix} 
-0.157351\\
   0.870170\\
   0 \\
   1.46598\\
  0.974368\\
   1.49816\\
 -0.420367\\
   1.35362\\
  0.790624\\
   1.49209
\end{pmatrix} ,
\end{equation}
and we get the returning cone, $C=\{v | Rv>0\}$, given by
\begin{equation}
\begin{pmatrix} 
                                        5.02913v_5 + 2.0v_6 + 0.970874v_7  \\
               0.970874v_1 + 22.8596v_5 - 7.47045v_7 + 5.02913v_9 + 2.0v_{10} \\
 - 7.47045v_1 + 5.02913v_3 +2.0v_4  + 104.878v_5 - 33.9565v_7 + 22.8596v_9  \\
   - 33.9565v_1 + 22.8596v_3 + 453.855v_5 - 150.760v_7 +2.0v_8 + 104.878v_9  \\
   - 150.760v_1 +2.0v_2 + 104.878v_3 + 1959.06v_5 - 651.314v_7 + 453.855v_9  \\
\end{pmatrix} > 0.
\end{equation}
It is easy to check that:
\begin{equation}
Rv^*=
\begin{pmatrix} 
 7.48842 \\
 32.2215 \\
 138.644 \\
 596.565 \\
 2566.93  
\end{pmatrix} > 0.
\label{Cone}
\end{equation}
It is worth noting that in~\eqref{eq:switchineq} or \eqref{eq:Rx>0}, we omit the denominators of the maps, since they are always positive and do not affect the inequality. If we include the denominators of the maps that make up $C$, the matrix $R$ changes, but only by a positive scaling in each row so that the inequality is equivalent, and we have

\begin{equation}
R(v^*)=
\begin{pmatrix} 
1.74034 \\
1.74034 \\
1.74034 \\
1.74034 \\
1.74034  
\end{pmatrix} > 0,
\label{Cone1}
\end{equation} 
where the equal positive values for each component of the inequality reflect the symmetry of the circuit.  

This establishes that there is a stable periodic orbit that follows this cycle, with period $\log(\lambda_{\max})=\log(1474.9579)=7.29638$.

The 5-fold symmetry of this circuit design with the chosen parameters means that we could also identify fixed points by following just two steps of the cycle from a starting wall, and then rotating variables backwards so that we are back on the original wall. A periodic orbit of the 10-cycle must have this type of symmetry, where after two steps we are at a permuted version of the original point, because if not, then there would be another, distinct dominant eigenvector (not a scalar multiple of the original one) for the same cycle, which would contradict the fact that this strictly dominant eigenvector here has a one-dimensional eigenspace according to the Perron-Frobenius theorem. This calculation is done in Appendix A.

\begin{figure}%
\begin{center}
\includegraphics[width=.8\columnwidth]{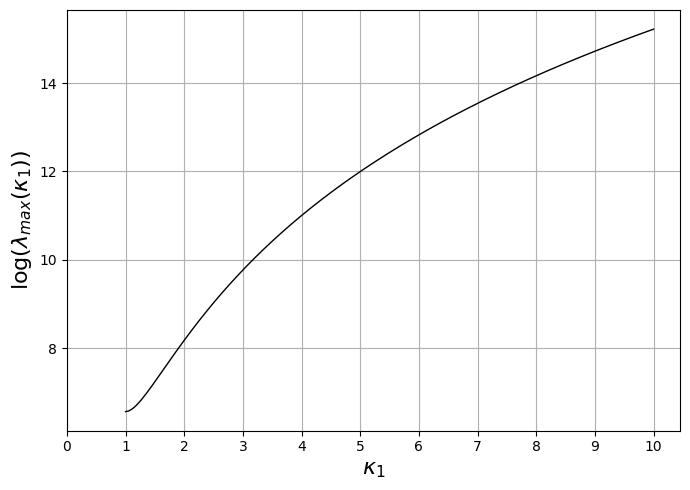}
\end{center}
\caption{The logarithm of the dominant eigenvalue $\lambda_{max}$ as a function of $\kappa_1$, for the $10$-dimensional model of the ring circuit.}
\label{figure13}
\end{figure}

\begin{figure}%
\begin{center}
\includegraphics[width=.8\columnwidth]{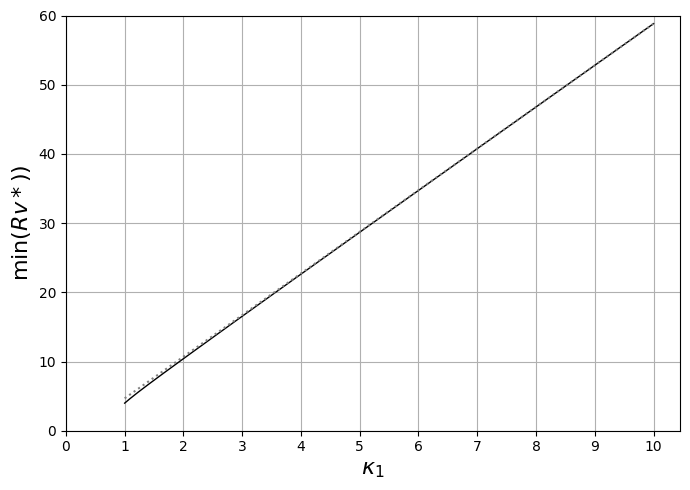}
\end{center}
\caption{The minimum component of the returning cone condition vector, $\min\{ R v^*\}$, at each fixed point as a function of $\kappa_1$, for the $10$-dimensional model of the ring circuit. A dotted straight line matching the slope of the graph at $\kappa_1\approx 10$ is shown to emphasise that $\min\{ R v^*\}$ does not itself vary linearly.}
\label{figure14}%
\end{figure}

Considering the full 10-step cycle and taking $\kappa_2=2$, a plot of the dominant eigenvalue of the matrix $B$ as a function of $\kappa_1$ is given in Figure~\ref{figure13} using a logarithmic scale for readability. This computation of the dominant eigenvalue for a series of values of $\kappa _1$ suggests that it is increasing with increasing $\kappa_1$ and that at $\kappa_1=1$ we already have $\lambda_{\max}>1$, satisfying condition \textsf{LC2} above for existence of a stable limit cycle. Of course, condition \textsf{LC3} is satisfied because this is the dominant eigenvalue. The minimum value of the vector $R v^*$ at the corresponding fixed point, $v^*$, is shown as a function of $\kappa_1$ in Figure~\ref{figure14}, and the fact that it is always positive here and increasing with $\kappa_1$ indicates that $v^*$ lies in the returning cone (condition $\textsf{LC1}$) so the system has a stable limit cycle for $\kappa_1>1$ (though this numerical demonstration is not a proof, and we have not attempted to verify that these graphs continue to increase for larger $\kappa_1$, though it seems likely).


\section{Analysis of the full $4n$-dimensional model of the ring circuit}\label{sec:4n}
\subsection{Analysis of a long cycle}
Now, consider the full $4n$-dimensional model~\eqref{eq:4n} with $n=5$ units and in each unit $i \in \{1,\ldots , 5\}$, $\kappa_{x_i}=\kappa_1$, $\kappa_{y_i}=\kappa_2$, $\kappa_{z_i}=\kappa_3$, $\kappa_{u_i}=\kappa_4$, and $\gamma_{x_i}=\gamma_{y_i}=\gamma_{z_i}=\gamma_{u_i}=\gamma=1$, as well as $\theta=\frac 12$ for all switching functions. If we translate thresholds to $0$, the system is

\begin{equation}\label{eq:4nv1}
\begin{array}{lcl}
\displaystyle \frac{dv_{4i-3}}{dt} & = & \kappa_1 (s^+(v_{4i-5})s^-(v_{4i})+s^-(v_{4i-5})s^+(v_{4i})- \frac 12 - v_{4i-3}\\[3mm]
\displaystyle \frac{dv_{4i-2}}{dt} & = & \kappa_2 s^-(v_{4i-3})- \frac 12 - v_{4i-2}\\[3mm]
\displaystyle \frac{dv_{4i-1}}{dt} & = & \kappa_3 s^-(v_{4i-2})- \frac 12 - v_{4i-1}\\[3mm]
\displaystyle \frac{dv_{4i}}{dt}   & = & \kappa_4 (1-s^-(v_{4i-1})s^-(v_{4i+3})- \frac 12 - v_{4i},
\end{array}
\end{equation}
where
$$
(v_{4i-3} \quad v_{4i-2} \quad v_{4i-1} \quad v_{4i} )^\top=\left(x_i -\frac 12 \quad  y_i -\frac 12 \quad  z_i-\frac 12  \quad u_i-\frac 12 \right)^{\top}, \quad i=1,\ldots, n\,.
$$

Our objective is to show that where irregular behaviour appears numerically in certain parameter intervals, we can analytically track stable periodic orbits in neighbouring intervals, identify the bifurcation point where the stable periodic orbit is lost, and demonstrate that the non-smooth bifurcation is of a type for which there is no stable periodic orbit (locally) on the other side.

Throughout this section, we will take (somewhat abritrarily) $\kappa_1=\kappa_2=1.06$ and $\kappa_3=\kappa_4$ (and still $\gamma=1,\, \theta=\frac 12$). Furthermore, we do all calculations in extended precision with $64$ decimal digits. The values in the matrix products after hundreds of steps become very large as do some eigenvalues of these matrices, and other eigenvalues are many orders of magnitude smaller, so extended precision is needed.

\begin{figure}[htb]%
\includegraphics[width=1.0\columnwidth]{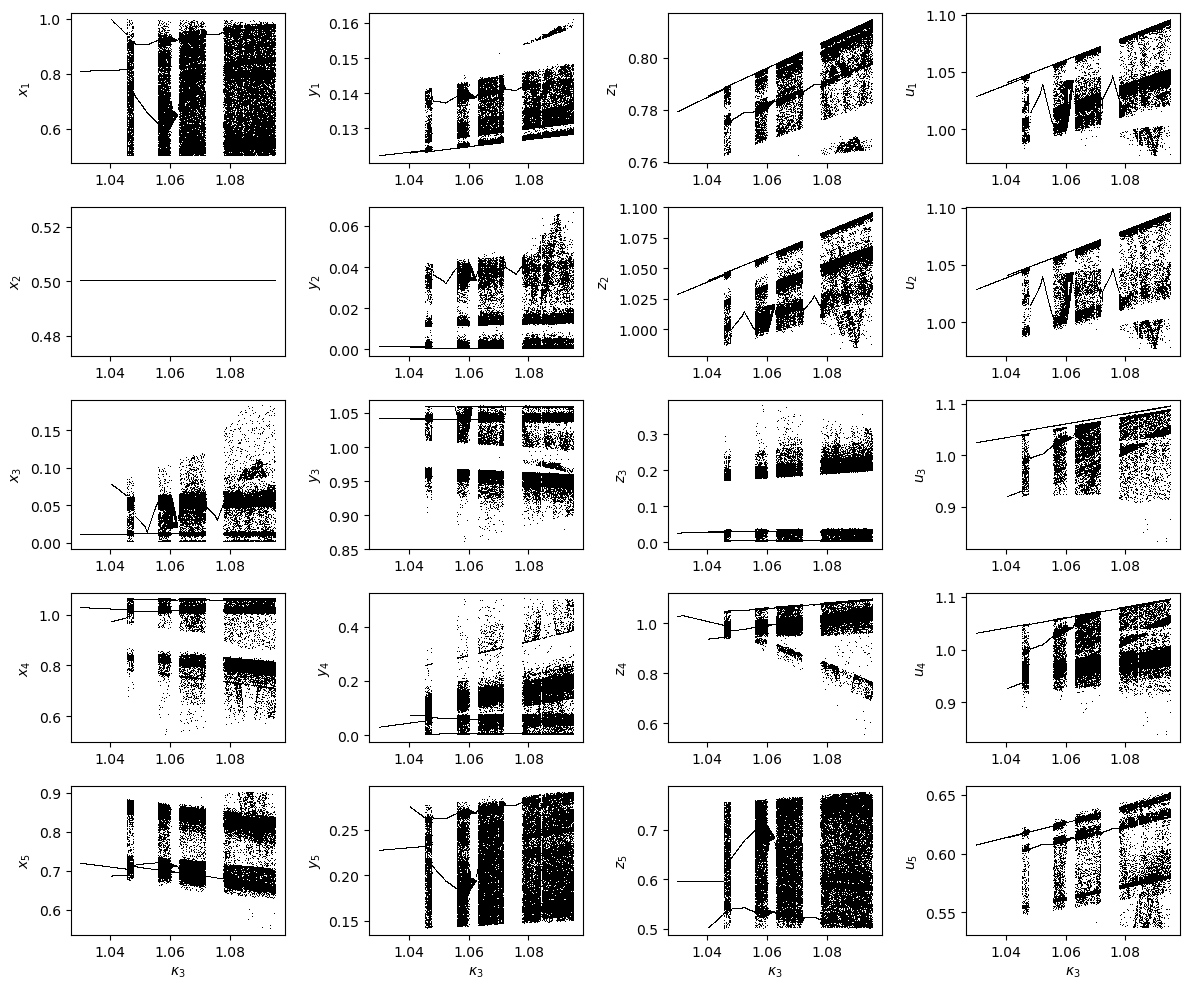}%
\caption{Bifurcation diagram for all coordinates, plotting the last $100$ points returning to the starting wall as a function of $\kappa_3$, using $30,000$-step trajectories, for the $20$-dimensional model of the ring circuit. The initial condition is as in the text; $\kappa_1=\kappa_2 =1.06, \kappa_4=\kappa_3, \gamma=1, \theta =\frac 12$.}%
\label{figure:allbifs}%
\end{figure}

\begin{figure}[htb]%
\centering
\includegraphics[width=.8\columnwidth]{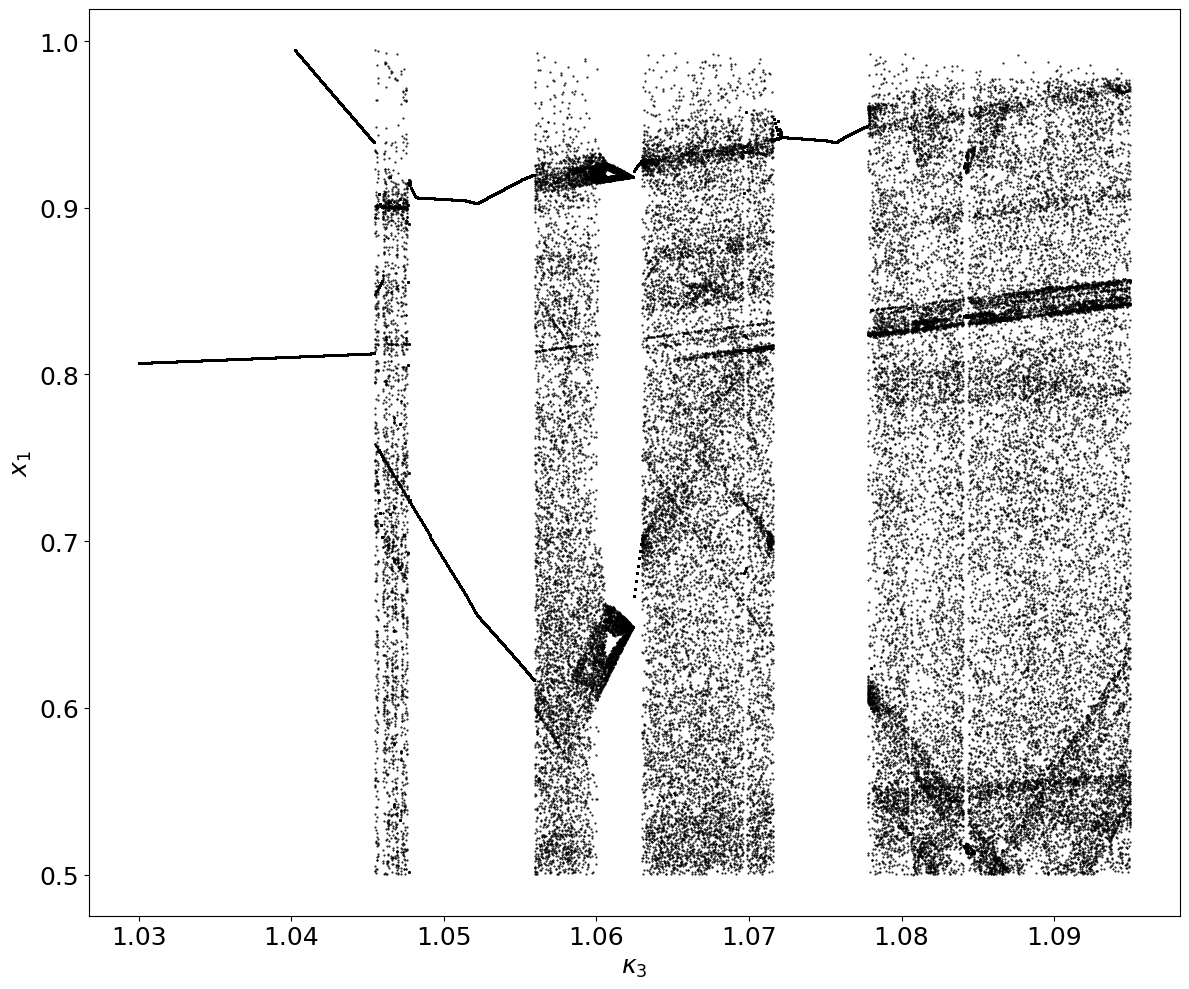}%
\caption{Expanded bifurcation diagram for the $x_1$ coordinate (taken from Figure~\ref{figure:allbifs}), plotting the last $100$ points returning to the starting wall as a function of $\kappa_3$.
}%
\label{fig:bif_103_1095_x1}%
\end{figure}

A bifurcation diagram as $\kappa_3$ varies is shown in Figures~\ref{figure:allbifs} and \ref{fig:bif_103_1095_x1}. It is constructed as follows. We start with $\kappa_3=1.03$ and from the initial condition in the original variables: 
$$x_0=\left(0.9     \quad   0 \quad     0.9   \quad     0.9    \quad    0.5   \quad       0  \quad   0.9 \quad       0.9 \quad            0 \quad    0.9    \quad      0  \quad    0.7\quad        0.8   \quad          0  \quad   0.8    \quad    0.8  \quad      0.8   \quad      0   \quad   0.8      \quad    0.8 \right)^{\top} ,$$
where $x_2=\frac 12$ is at its threshold value. The starting wall can be represented in terms of boxes as 
$$ \left(1 \quad 0 \quad 1 \quad 1 \quad 0.5 \quad 0 \quad 1 \quad 1 \quad 0 \quad 1 \quad 0 \quad 1 \quad 1 \quad 0 \quad 1 \quad 1 \quad 1 \quad 0 \quad 1 \quad 1 \right).$$
For each of $1000$ equally-spaced values of $\kappa_3$ in the interval $[1.03,1.095]$ (including the endpoints), we compute a trajectory of 30000 steps (wall crossings) explicitly from the mappings from wall to wall (given by Equations~\eqref{eq:exitmap},\eqref{eq:Bk},\eqref{eq:psik} in translated coordinates), and plot each original coordinate of the last $100$ points of return to the starting wall. For the initial condition at the next value of $\kappa_3$ (moving to the right), we start from the last return point in the starting wall for the previous value of $\kappa_3$. 

\begin{figure}[htb]
\begin{center}
\includegraphics[width=0.8\columnwidth]{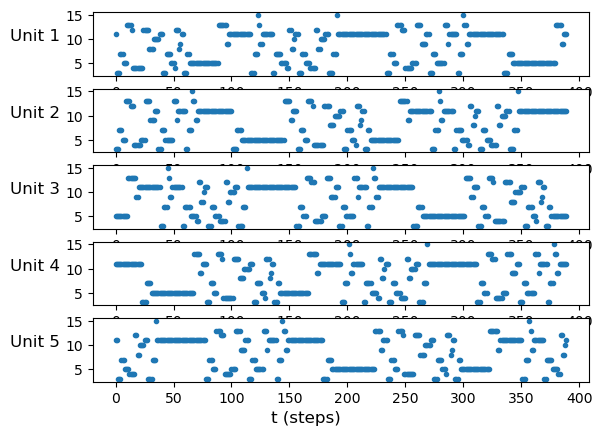}
\end{center}
\caption{The sequence of states followed by the cycle of length 390, represented as the decimal equivalent of the binary sequence for each unit. Cycle steps are indexed by the horizontal axis. Note that the 5 graphs for the 5 units are identical under a shift of 156 steps (2/5 of the cycle) between successive units.}
\label{fig:390stateseq}
\end{figure}

Now, to focus on a particular parameter value, 
take $\kappa_3=\kappa_4=1.055$, and start from initial condition:
$$ x_0= \left(1 \quad  0 \quad     1 \quad    1 \quad   0.5\quad  0\quad     1\quad    1\quad         0\quad     1 \quad        0 \quad     1\quad    1 \quad        0  \quad   1  \quad  1\quad    1    \quad     0  \quad  1 \quad   1 \right )^{\top}.$$
Computing a sequence of wall crossings, the trajectory appears to converge to a cycle of boxes with $390$ transition steps before repeating. The sequence of states is depicted in Figure~\ref{fig:390stateseq}, where the state of each unit is represented as a decimal equivalent (0 to 15) of the binary state (0000 to 1111).

Another way to represent the cycle is by the sequence of switching variables, {\em i.e.}, the sequence of indices of variables that switch in temporal order, which for this 390-step cycle is given by: \\

\noindent
\begin{tabular}{rrrrrrrrrrrrrrrrrrrrrrrrrr}
 1,& 17,&  6,&  2,& 18,&  7,&  3,& 19,&  5,&  1,&  9,& 20,&  8,&  4,&  1,&  5,& 17,& 10,& 18,& 11,& 19,&  8,& 13,&  1,& 20,&  5, \\
14,& 17,&  2,&  6,& 15,& 18,& 3,&  7,& 17,& 18,&  5,&  4,&  9,&  6,&  1,& 10,&  7,&  2,&  9,& 11,& 10,&  3,& 11,&  5,&  4,&  1, \\
6,&  7,&  2,&  4,&  3,&  9,&  1,& 10,&  5,&  2,&  6,&  3,& 11,&  5,&  7,& 13,&  6,& 12,&  7,&  9,& 14,& 10,& 15,& 11,& 12,& 17, \\
13,&  9,& 18,& 14,& 10,& 19,& 15,& 11,& 17,& 13,&  1,& 12,& 20,& 16,& 13,& 17,&  9,&  2,& 10,&  3,& 11,& 20,&  5,& 13,& 12,& 17, \\
6,&  9,& 14,& 18,&  7,& 10,& 15,& 19,&  9,& 10,& 17,& 16,&  1,& 18,& 13,&  2,& 19,& 14,&  1,&  3,&  2,& 15,&  3,& 17,& 16,& 13, \\
18,& 19,& 14,& 16,& 15,&  1,& 13,&  2,& 17,& 14,& 18,& 15,&  3,& 17,& 19,&  5,& 18,&  4,& 19,&  1,&  6,&  2,&  7,&  3,&  4,&  9,  \\
5,&  1,& 10,&  6,&  2,& 11,&  7,&  3,&  9,&  5,& 13,&  4,& 12,&  8,&  5,&  9,&  1,& 14,&  2,& 15,&  3,& 12,& 17,&  5,&  4,&  9, \\
18,&  1,&  6,& 10,& 19,&  2,&  7,& 11,&  1,&  2,&  9,&  8,& 13,& 10,&  5,& 14,& 11,&  6,& 13,& 15,& 14,&  7,& 15,&  9,&  8,&  5, \\
 10,& 11,&  6,&  8,&  7,& 13,&  5,& 14,&  9,&  6,& 10,&  7,& 15,&  9,& 11,& 17,& 10,& 16,& 11,& 13,& 18,& 14,& 19,& 15,& 16,& 1, \\
 17,& 13,&  2,& 18,& 14,&  3,& 19,& 15,&  1,& 17,&  5,& 16,&  4,& 20,& 17,&  1,& 13,&  6,& 14,&  7,& 15,&  4,& 9,& 17,& 16,&  1, \\
  10,& 13,& 18,&  2,& 11,& 14,& 19,&  3,& 13,& 14,&  1,& 20,&  5,&  2,& 17,&  6,&  3,& 18,&  5,&  7,&  6,& 19,&  7,&  1,& 20,& 17, \\
 2,&  3,& 18,& 20,& 19,&  5,& 17,&  6,&  1,& 18,&  2,& 19,&  7,&  1,&  3,&  9,&  2,&  8,&  3,&  5,& 10,&  6,& 11,&  7,&  8,& 13,  \\
 9,&  5,& 14,& 10,&  6,& 15,& 11,&  7,& 13,&  9,& 17,&  8,& 16,& 12,&  9,& 13,&  5,& 18,&  6,& 19,&  7,& 16,&  1,&  9,&  8,& 13, \\
  2,&  5,& 10,& 14,&  3,&  6,& 11,& 15,&  5,&  6,& 13,& 12,& 17,& 14,&  9,& 18,& 15,& 10,& 17,& 19,& 18,& 11,& 19,& 13,& 12,&  9, \\
  14,& 15,& 10,& 12,& 11,& 17,&  9,& 18,& 13,& 10,& 14,& 11,& 19,& 13,& 15,&  1,& 14,& 20,& 15,& 17,& 2,& 18,&  3,& 19,& 20,&  5. 
\end{tabular}
\ \\

\begin{figure}[htb]
\includegraphics[width=.99\columnwidth]{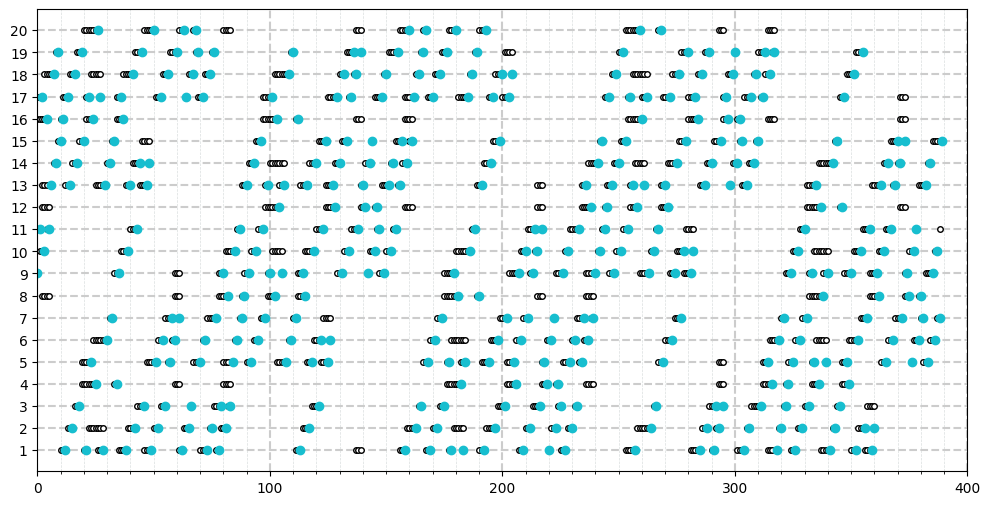}
\caption{The sequence of switching variables (solid circles) and alternate switching variables (small open circles) at each step in the cycle of length 390. The cycle steps are indexed by the horizontal axis. The variables are indexed by the vertical axis. Note that the generally forward propagation of perturbations around the cycle is visible by the trend toward upward diagonal patterns in the plot. Note also that when one variable is one of several alternatives for a few steps, once it actually switches on the cycle, it ceases to be an alternate switching variable for a while (closed circles terminate a series of open circles).}
\label{fig:390alternateswitchseq}
\end{figure}

This sequence is represented graphically in Figure~\ref{fig:390alternateswitchseq}. The boxes visited along the cycle could be reconstructed from the initial wall and the sequence of switching variables.


For this cycle, a sufficiently long trajectory appears to converge to a periodic orbit starting and returning to a point on the starting wall at 
\begin{equation}
x^*= 
{\tiny 
\begin{pmatrix} 
0.6264\\    0.1241  \\  0.7932  \\  1.0550  \\  0.5000 \\   0.0001  \\  1.0550  \\  1.0550   \\ 0.0117 \\   1.0396 \\   0.0298  \\  1.0147 \\  1.0122\\    0.0604 \\   0.9798  \\  1.0214
\\
 0.7183   \\ 0.1865  \\  0.6927  \\  0.6226

\end{pmatrix} }\,.
\label{fixed55numeric}
\end{equation}

Starting at this point, the trajectory returns to the starting wall after $292$ steps but at a different point, 
\begin{equation}
x^{**}= 
{\tiny
\begin{pmatrix} 
0.9164   \\
 0.1385   \\ 0.7788   \\ 1.0087  \\  0.500 0\\ 0.0378  \\
   1.0017    \\1.0093   \\ 0.0466    \\1.0586  \\  0.0026  \\  1.0550 \\   1.0562   \\ 0.0040  \\  1.0522 \\   1.0541\\
    0.7021 \\   0.2658  \\  0.5340\\    0.6075
\end{pmatrix} }\,,
\label{fixed551}
\end{equation}
then again after $390$ steps at the starting point. These two points can be seen in the bifurcation diagram (Figures~\ref{figure:allbifs}, \ref{fig:bif_103_1095_x1}) at $\kappa_3=1.055$.
Thus, there is another cycle of length 390 from the same starting wall, starting from step $292$ in the first cycle above and following the same sequence of steps from there. 
This is therefore a phase-shifted version of the original cycle.

The existence and stability of the periodic orbits through these cycles can be confirmed analytically. For the first cycle above, translating the thresholds to the origin, the solution at each step around the cycle 
is calculated explicitly according to
\begin{equation}
v^{(k+1)}=\frac{B^{(k)} v^{(k)}}{1 + \psi^{(i)}{}^{\top} v^{(k)}}
\label{eq:PMap2}
\end{equation}
as described in Section~\ref{sec:BestCircuit},
and the dominant eigenvalue is
$$\tiny \lambda_{max} = 232878744650361409031330479377078.9569064920993228397022$$
and the corresponding fixed point of the cycle map, a scaled version of the dominant eigenvector, $w_{\mbox{max}}$, is given by
\begin{equation}
v^*= (\lambda_{\max}-1)\frac{w_{\max}}{\psi_{390}^{\top} w_{\max}} = 
{\tiny
\begin{pmatrix} 
  0.12635206507785982451\\
 -0.37589167932511592714\\
  0.29317436517128393701\\
  0.55499998870162586423\\
  0.0 \\
 -0.49993394928697622404\\
  0.55495311595634358236\\
  0.55498412965520054305\\
 -0.48826475566603278518\\
  0.53956448330312469355\\
 -0.47021020024632053415\\
  0.51468565657952449326\\
  0.51216384275672264934\\
 -0.43961212248543262666\\
  0.47978313540285016477\\
  0.52139205775579350582\\
  0.21831868247200336912\\
 -0.31346679093389828349\\
  0.19271549384083334821\\
   0.1225754831586405685
\end{pmatrix} },
\label{fixed_055a}
\end{equation}
After moving the thresholds back to $\frac{1}{2}$, we get in the original coordinates:
\begin{equation}
x^*=v^*+\frac{1}{2}=
{\tiny 
\begin{pmatrix} 
  0.6263520651 \\
     0.1241083207\\
     0.7931743652\\
      1.054999989\\
      0.5 \\
 0.00006605071302\\
      1.054953116\\
       1.05498413\\
    0.01173524433\\
      1.039564483\\
    0.02978979975\\
      1.014685657\\
      1.012163843\\
    0.06038787751\\
     0.9797831354\\
      1.021392058\\
     0.7183186825\\
     0.1865332091\\
     0.6927154938\\
     0.6225754832
\end{pmatrix}} ,
\end{equation}
which is exactly the point found numerically in Equation~\eqref{fixed55numeric}.

The sequence of alternative branching variables at each step of the cycle can be identified by comparing the focal point coordinate with the current state, and noting sign differences. Thus, for example, at the fist step, where variable $1$ switches (i.e., $v_1$), variables $6$ and $17$ are alternate exit variables. A tabulation of all these alternate exit variables shows that at every step there exists at least one alternative branching variable, and there are $1025$ alternate exit variables in all. The sequence of alternate exit variables, as well as actual exit variables, at each step of the cycle is shown in Figure~\ref{fig:390alternateswitchseq}.



The conditions (inequalities) defining the returning cone for the cycle can be calculated explicitly. For example, at the first variable, the alternate exit variable $v_6$ leads to the inequality on the starting wall: 
$$ - 2 v_1 - 1.78571v_6>0 $$
and the alternate exit variable $v_{17}$ at step 390 of the cycle leads to the inequality on the starting wall:
\begin{eqnarray*}
- 25822063140750828468335157166288.459v_1 + 10244087486334445609316533301700.594v_2 \\
- 5048987865199785345080290707525.2272v_3 - 6003894224261788474521775143052.1282v_4 \\
- 3788234465630294647126765140035.6902v_5 - 36562467287687952493789450584124.494v_6 \\
+ 19750099092453495912887605783326.905v_7 - 7730940093506221615893059258803.0149v_8 \\
- 17143383946513774463186825415276.196v_9 + 9117361524360924472538558529432.4759v_{10} \\
- 4336650919343164918293141601072.6989v_{11} + 393855799605144203478675392.85434825v_{12} \\
- 3622761307706059225852617991.6447493v_{13} + 1253401371906587663553710232.768513v_{14} \\
- 396886767161023367506253510.2944945v_{15}  - 1997955097008048544698561.352860036v_{16} \\
+ 165996981250004374308755161542872.26v_{17} - 82498882624037904627256987475157.971v_{18} \\
 +  46286890768707657272999418316394.56v_{19} + 18869074277779440678866011836229.869v_{20} >0 .  
\end{eqnarray*}
The latter inequality has large coefficients because they accumulate in computing the maps back around the cycle to the starting wall. Extended precision calculations are again required. It is worth noting that any such inequality can be scaled by an arbitrary positive number and thus be brought down into a reasonable range. If the left hand side is close to zero, though, it is sensitive anyway and precision is needed.
The returning cone is defined by $1025$ such conditions corresponding to the alternate exit variables. 

Evaluating these 1025 inequalities defining the returning cone at the fixed point in \eqref{fixed_055a} we find that all conditions are satisfied ($Rv^*>0$). Therefore, the fixed point is in the domain of definition for the cycle and corresponds to a stable periodic orbit of the system.

For the phase-shifted version of the cycle, the sequence of switching variables and alternate switching variables are the same apart from the phase shift, but the returning cone in the starting wall will of course be different. The condition $Rv^*>0$ is, of course, also satisfied there.

We conclude that even if the dimension is high (20) and the number of steps in the cycle is large (more than $300$ steps), the calculations of the cycle map and the returning cone conditions are still possible, albeit with extended precision calculations. 

\subsection{Multi-stability and bifurcations of periodic orbits}

\begin{figure}
\centering
  \includegraphics[width=.75\columnwidth]{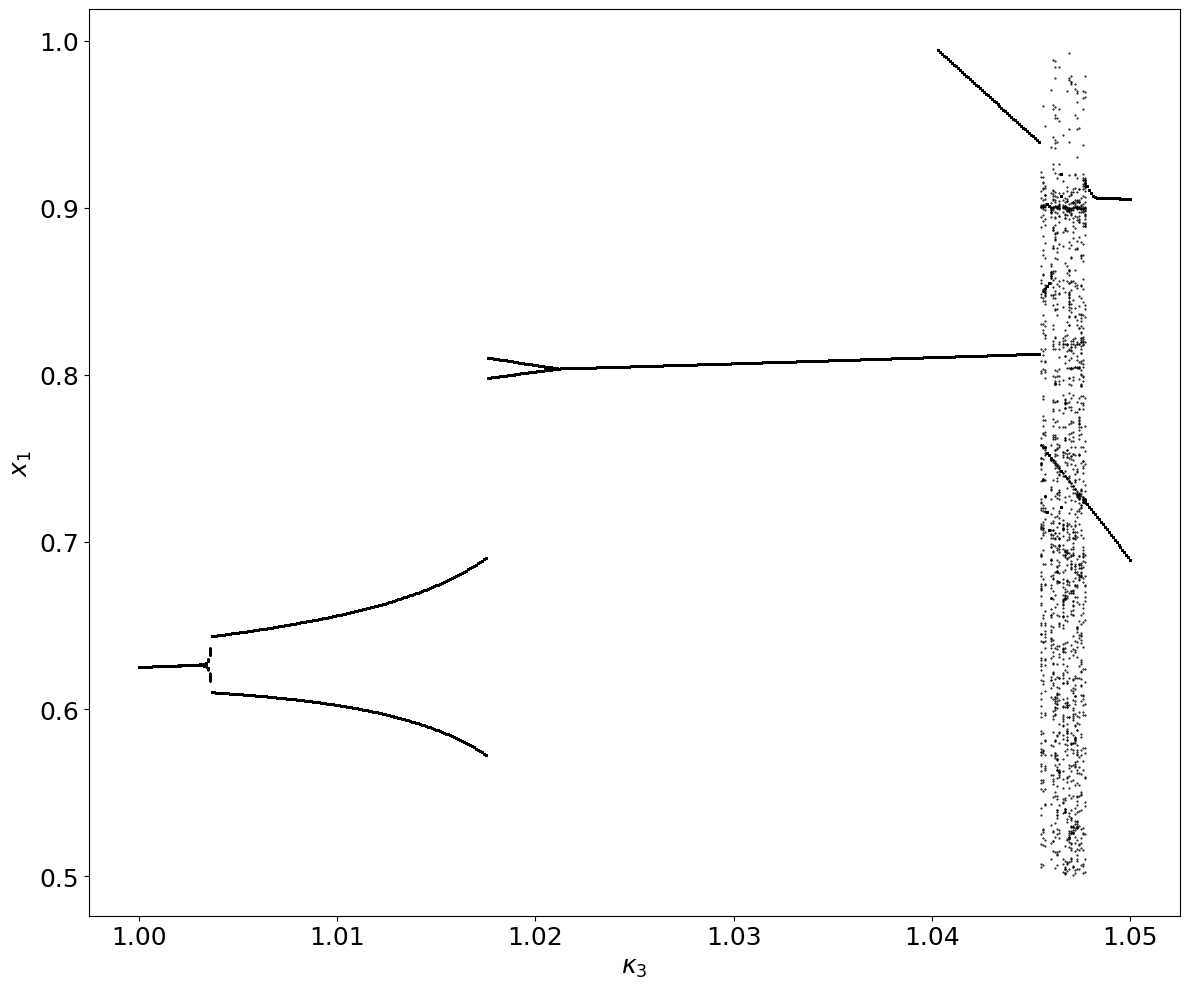}%
\caption{Bifurcation diagram for parameter value $1.00<\kappa_3<1.05$. Note the period-doubling bifurcation as $\kappa_3$ decreases through $\approx 1.02118$. The bifurcation as $\kappa_3$ increases through $\approx 1.00326$ is actually a clean discontinuous jump from a length $170$ cycle to a different cycle of length $340$ at a cycle destabilizing bifurcation. The additional points that appear in the diagram are transients. A very long integration is needed to move past these transients.}
\label{fig:bif_100_105}%
\end{figure}

Without giving all the details, we mention first that there is an example of a {\bf CD(a)} bifurcation in this system at $\kappa_3\approx 1.00327$. From $\kappa_3=1.000$ until $\kappa_3=1.00326$ there is a stable cycle of length $170$. At $\kappa_3=1.00326$, the largest two eigenvalues of the cycle map matrix are 
\begin{eqnarray*}
    \lambda_1 &=& \,\,\,\,\,212451617877993.014871 \quad\mbox{and} \\
    \lambda_2 &=& -212451309163520.086517\,. 
\end{eqnarray*}
The largest two eigenvalues of the same cycle at $\kappa_3=1.00327$ are
\begin{eqnarray*}
\lambda_1 &=& -212462503488602.451419 \quad\mbox{and} \\
\lambda_2 &=& \,\,\,\,\,212422586198129.303063\,.
\end{eqnarray*}
The top two eigenvalues have swapped places, and the eigenvector associated with the dominant one at $\kappa_3=1.00326$ is associated with the second-largest eigenvalue at $\kappa_3=1.00327$. Thus, the periodic orbit through the fixed point of the cycle map on this eigenvector is no longer stable. Trajectories starting near the now unstable fixed point of the $170$-cycle map now converge to another cycle of length $340$. See Figure~\ref{fig:bif_100_105}. Note that this is not a period-doubling bifurcation. There is no periodic orbit after the bifurcation, locally, that is directly related to the stable periodic orbit before the bifurcation. The new stable periodic orbit is distinct.

Because these bifurcation diagrams track a single attractor for each value of $\kappa_3$, they do not indicate the possible presence of multiple stable attractors at the same value of $\kappa_3$. Instances of multistability will be described below.

 Consider now the part of the bifurcation diagram in Figure~\ref{fig:bif_103_1095_x1} to the right of $\kappa_3=1.0745$ leading to what appears to be a bifurcation to chaos. We will still fix $\kappa_1=\kappa_2=1.06$ and $\kappa_3=\kappa_4$ throughout. 
 
 We first look at the situation where $\kappa_3=1.0775$. Our initial condition is:
$$x_0= \left(1 \quad  0 \quad  1 \quad 1 \quad   0.5\quad  0\quad  1\quad 1\quad   0\quad  1 \quad  0 \quad  1\quad 1 \quad  0  \quad   1  \quad  1\quad 1 \quad  0  \quad  1 \quad   1 \right )^{\top}$$
Numerically, we identify a cycle with $96$ steps before returning to the same point on the starting wall.

The sequence of transition variable indices is given by: \\
\begin{tabular}{rrrrrrrrrrrrrrrrrrrrrrrr}
17,& 1,& 6,&   18,&   2,&  19,&  7,&  1,& 3,&   9,& 2,& 8,&  3,& 5,&  10,& 6,&   11,&  7,&  8,& 13,&  9,&  5,& 14,& 10, \\
6,& 11,& 15,&  7,& 13,&  9,& 17,& 8,&  12,& 16,&   9,&   13,& 5,&  18,&   6,&   19,& 7,& 16,&  1,&  9,& 8,& 13,&  2,&  5, \\
 10,&   14,& 3,& 6,& 11,& 15,& 5,& 6,& 13,&   12,&   17,&   14,&   9,&   18,& 15,&   10,& 17,&   19,& 18,&   11,&   19,& 13,& 12,& 14, \\
  9,&   15,&   10,& 12,&  11,& 17,& 9,&   18,& 13,&   10,&   14,& 11,& 19,&   13,&   14,&   1,& 20,& 17,&  2,&  3,& 18,& 20,& 19,& 5.\\[2mm]
\end{tabular}

\begin{figure}[htb]
\includegraphics[width=.99\columnwidth]{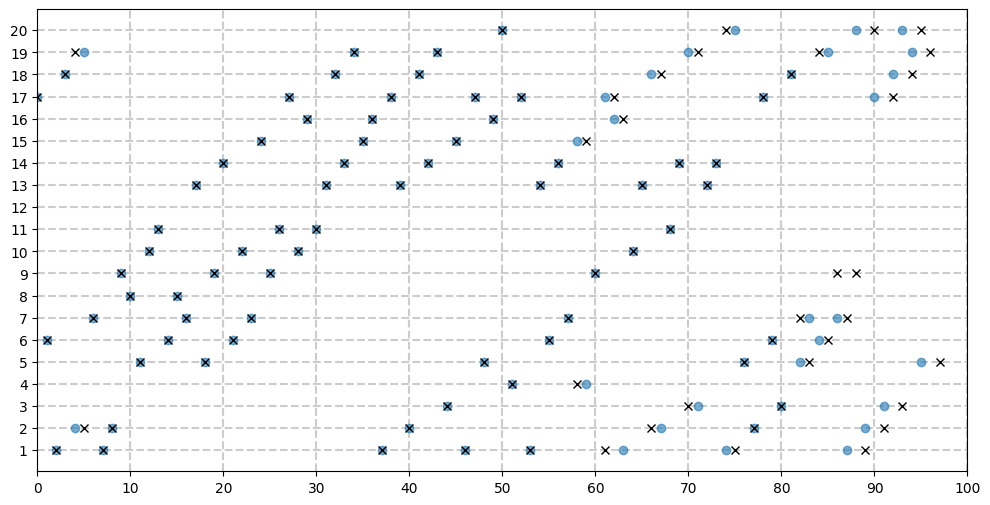}
\caption{The sequence of switching variables at each step in the cycle of length 96 at $\kappa=1.0775$ are shown by circles. The sequence of switching variables at each step in the cycle of length 98 at $\kappa=1.07779359$ are shown by crosses. The cycle steps are indexed by the horizontal axis. The variables are indexed by the vertical axis.}
\label{fig:96switchseq}
\end{figure}   

As before, we can compute the cycle map for this 96-step cycle and the dominant eigenvector, as well as the 237 inequalities defining the returning cone, and thereby confirm that the fixed point on the dominant eigenvector lies in the returning cone, and thus corresponds to a stable periodic orbit for the circuit.

Starting from another initial point in the same box:
$$ \begin{array}{cccccccccc}
\tilde{x}_0=(0.9480 &   0.1415  &   0.7895 &   1.0276 &   0.5000 &   0.0398 &   1.0193 &   1.0282 &   0.0492 &   1.0585 \\
0.0029 &   1.0775 &   1.0557 &   0.0048 &   1.0731 &   1.0765 &   0.6786 &   0.2789 &   0.5172& 0.6209 )^{\top}\end{array}$$  
we find a different 96-step cycle. The sequence of transitions is given by: \\
\begin{tabular}{rrrrrrrrrrrrrrrrrrrrrrrr}
 17,&  6,&  1,& 18,&  2,& 19,&  7,&  1,&  2,&  9,&  8,&  5,& 10,& 11,&  6,&  8,&  7,& 13,&  5,&  9,& 14,&  6,& 10,&  7,  \\
15,&  9,& 11,& 17,& 10,& 16,& 11,& 13,& 18,& 14,& 19,& 15,& 16,&  1,& 17,& 13,&  2,& 18,& 14,& 19,&  3,& 15,&  1,&   17,   \\
5,& 16,& 20,&  4,&   17,&  1,& 13,&  6,&   14,& 7,&   15,&   4,&   9,&   17,& 16,&  1,&  10,& 13,&   18,&   2,&   11,&  14,&   19,&  3,   \\
13,& 14,&  1,&   20,&   5,&  2,&  17,& 6,&  3,& 18,& 5,&  7,&  6,& 19,& 7,& 1,&   20,& 2,&   17,& 3,& 18,&   20,& 19,&  5. 
\end{tabular}

\ \\
This switching sequence is shown graphically by the circles in Figure~\ref{fig:96switchseq}.

The fixed point
\begin{equation}
x^*= 
{\tiny
\begin{pmatrix} 
  0.9479796977 \\  0.1414646605 \\   0.7894549447 \\   1.027604556 \\   0.5 \\   0.03981459955 \\   1.019258527 \\   1.028220836 \\   0.04917847230 \\   1.058514338 \\   0.00285657361 \\   1.077498354 \\   1.055687772 \\   0.00476965298 \\   1.073140827 \\   1.076454397 \\   0.6786124605 \\   0.2788804811 \\   0.5172253354 \\   0.6208612346 
\end{pmatrix} ,
}
\label{fixed96numericb}
\end{equation}
can again be computed analytically from the cycle map, the dominant eigenvector, and the returning cone, and it can be confirmed that it contains the dominant eigenvector, and thus this fixed point.

This establishes that there are at least two stable limit cycles from this starting wall, following somewhat different 96-step cycles. In fact, other cycles can be found for other initial conditions on the same wall (not shown here). 
The second cycle above is actually equivalent to the first one but rotated $3$ units forward (similar to what happened in the $2n$ model), or $2$ units backward, and shifting the second sequence $18$ steps backward. This is because of the symmetry of the system. 
Both of these periodic orbits must exist on the same interval of parameter values and they are lost at the same bifurcation point. 
We conclude that multistability of cycles exists in this system.  

\begin{figure}[htb!]%
\begin{center}	
\includegraphics[width=.8\columnwidth]{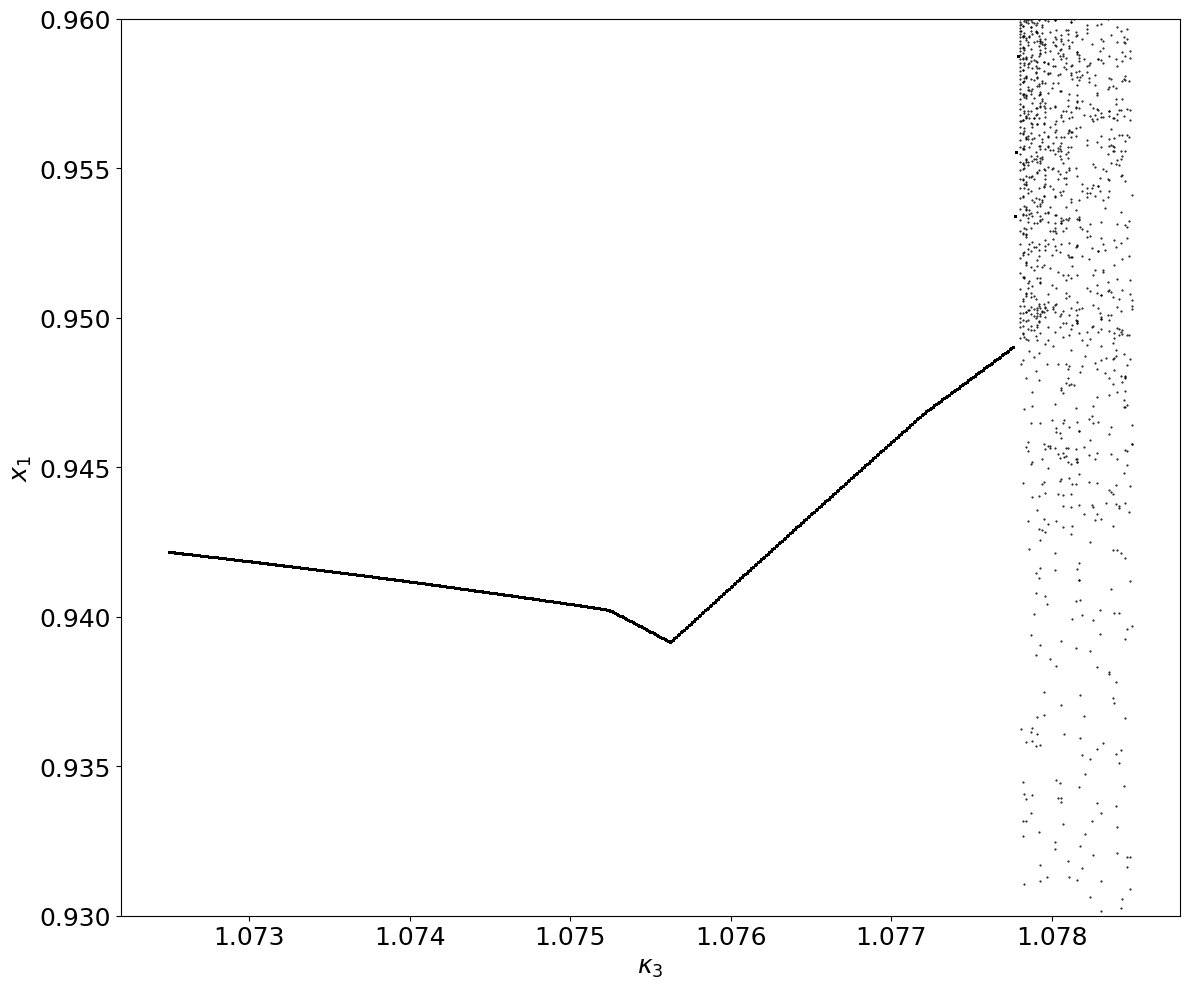}%
\caption{Bifurcation diagram over the interval $1.0725\le\kappa_3\le 1.0785$. Again, the $x_1$ coordinate of the last $100$ points returning to the starting wall are plotted as a function of $\kappa_3$. }
\label{fig:bif_10725_10825}
\end{center}
\end{figure}

A bifurcation diagram on a narrower interval, using $\tilde{x}_0$ as first initial condition 
and starting at $\kappa_3=1.0725$ is shown in Figure~\ref{fig:bif_10725_10825}. Variable $x_1$ is plotted on the same starting wall as before, 
$$ \left(1 \quad 0 \quad 1 \quad 1 \quad 0.5 \quad 0 \quad 1 \quad 1 \quad 0 \quad 1 \quad 0 \quad 1 \quad 1 \quad 0 \quad 1 \quad 1 \quad 1 \quad 0 \quad 1 \quad 1 \right),$$
and again for each increment of $\kappa_3$ we start from the last point reached in the same wall at the previous value of $\kappa_3$.


It is clear from Figure~\ref{fig:bif_10725_10825} that as the bifurcation parameter increases the system switches between a number of different cycles before it becomes chaotic. When a cycle is lost then another stable cycle becomes available, until there seem to be no more. It is possible to prove the existence and stability of these cycles and to track the values of $\kappa_3$ at which each stable cycle is lost. We can also identify the types of non-smooth bifurcations that occur at these points, and thus, what cycles, stable or unstable, must appear after the bifurcation, if any.

In the interval $1.0725<\kappa_3<1.07775$ (see Fig.~\ref{fig:bif_10725_10825}), a sequence of cycles, some of length $96$, others of length $98$, can be tracked in this way. In each case, cycles are lost through {\bf DS(b)} bifurcations, for which the change in cycles is $A\to B$. Thus, in each case, one cycle is continuously transformed into a similar one, through a slightly different sequence of boxes. Details for this interval are omitted, as we choose to focus on the final interval leading up to the transition to chaos.



\begin{figure}[htb!]%
\begin{center}	
\includegraphics[width=.8\columnwidth]{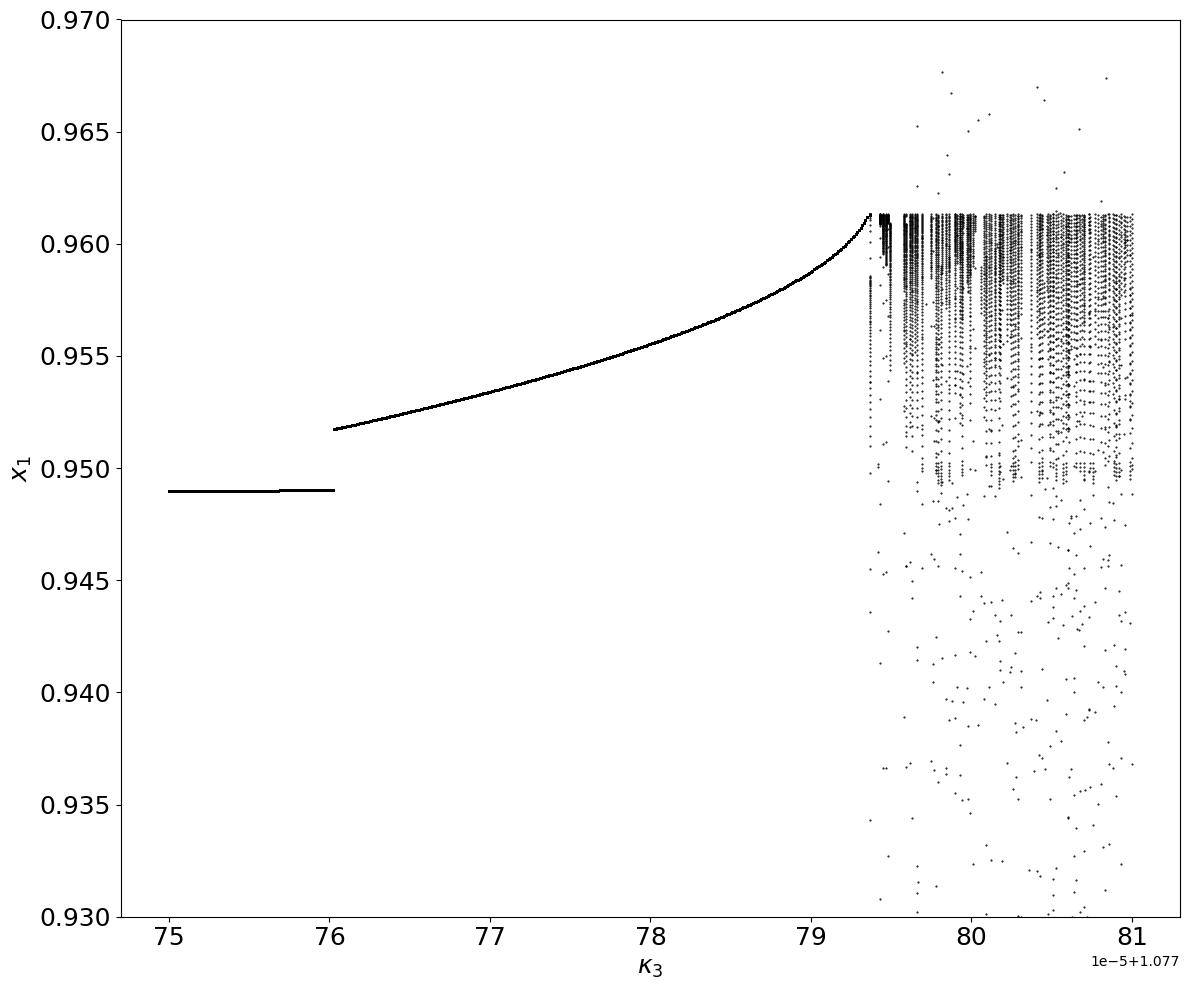}
\caption{Zoomed-in version of the previous bifurcation diagram, showing the sequence of bifurcations leading up to the transition to chaos. At each parameter value an integration of $60,000$ steps was made and the $x_1$ values of the last $100$ crossings of the starting wall plotted. In the interval $\left[ 1.07776030, 1.07779359\right]$ there exists a series of cycles of length $98$.}
\label{fig:bif_107775_107781}
\end{center}
\end{figure}

We can track bifurcations analytically as follows. First, we numerically identify a cycle to start from. Given a starting parameter value, we compute a trajectory from the wall-to-wall transition maps, and identify a cycle of boxes that are repeated periodically. We then check that there is a periodic orbit through this cycle of boxes by calculating the cycle map from a chosen starting wall, computing its returning cone, and the eigenvalues and eigenvectors of the matrix in the map. Since the initial numerical simulation converged on this cycle, we expect a periodic orbit to exist and to be stable (the dominant eigenvector must lie in the returning cone). 

Next we loop through increasing values of the parameter, incrementing at each step. At each new parameter value, we recalculate the map, eigenvalues, eigenvectors, and returning cone conditions for the cycle identified at the previous step, but using the new parameter value. If the eigenvector corresponding to the fixed point of the cycle (at the previous step) still lies in the cycle's returning cone and is still dominant, then we have the same stable cycle as at the last step. 

If not, then we attempt to find a stable periodic orbit beyond the bifurcation value, by one of several means.

If there is a swap in eigenvalues (the one associated with the fixed point we have been tracking becomes unstable or complex), or the dominant eigenvalue falls below $1$, then we flag the bifurcation type, and since we have lost the cycle, we compute a new trajectory (starting at the fixed point just before the bifurcation). 

If instead the eigenvector no longer lies in the returning cone (one or more of the conditions have become negative), then we have been through a double-switching bifurcation. The type can be identified by looking at the phase plane in two variables: the alternative switching variable that triggers the violation of a returning cone condition, and the one that switches at that point on the original cycle - {\em i.e.}, the two variables that switch simultaneously at some point on the cycle exactly at the bifurcation value. The focal points determine the change in switching variables between the original cycle and the one that is generated after the double switching. The eigenvalues of the maps for these two cycles then determine the type of double-switching bifurcation and the existence and stability of cycles on either side of the bifurcation, by Proposition~\ref{prop:DSbif}. 

If at this point we have a bifurcation of type $A\to B$ or $A\to b,AB$, then we have identified a stable periodic orbit beyond the bifurcation, and the parameter value can be incremented again, starting with this new stable cycle. If not, then we have no stable periodic orbit beyond the bifurcation, and we have to compute a new trajectory at this new parameter value (starting at the fixed point just before the bifurcation). 

If above we had to compute a new trajectory, then if this finds a stable periodic orbit, we continue from there. If not, we stop, and chaos is suspected.

Figure~\ref{fig:bif_107775_107781} zooms in on the final interval leading up to the transition to chaos.
Starting at $\kappa_3=1.0777500$ and the initial point $\tilde{x}_0$ shown earlier
to start on the branch of the length-96 cycle found above, and then incrementing $\kappa_3$ in steps of $10^{-7}$ we find that a periodic orbit through the same cycle exists and is stable up to $\kappa_3=1.0777602$, but as $\kappa_3>1.0777603$ it is lost through a double-switching bifurcation of type {\bf DS(c)}, which implies $A,b\to \emptyset$. Thus, there is no stable cycle (locally) beyond the bifurcation. However, computing a new trajectory at this value of $\kappa_3$, we find a new stable periodic orbit through a cycle of length 98. 

Continuing the incrementing of $\kappa_3$, we find that multiple bifurcations occur over a short interval, combined with the likely co-existence of several closely related cycles. This makes the analytical tracking bifurcations difficult in this narrow region.
Increasing $\kappa_3$ further, we find that at $\kappa_3=1.0777760$ the same 98-cycle as $\kappa_3=1.0777603$ is reached by time-stepping. We use it to track its branch as $\kappa_3$ goes through  $10^{-7}$ increments.  Immediately, at $\kappa_3=1.0777761$ there is a {\bf DS(b)} bifurcation, which implies $A\to B$. Thus, there is a stable periodic orbit through a new cycle, again of length 98. It differs from the 98-cycle found at $1.0777603$.

As we proceed to increase $\kappa_3$, we go through several more {\bf DS(b)} bifurcations, at or just before $\kappa_3= 1.0777821, 1.0777906$, and $1.0777919$, and the last cycle continues until $\kappa_3=1.0777935$. Then, finally, at $\kappa_3=1.0777936$ we lose this last cycle in a {\bf DS(c)} bifurcation. The information on this sequence of bifurcations is captured in Table~\ref{tab:bifseq}. We analyze the final bifurcation to chaos in more detail in the next subsection. 
\begin{table}[h]
\centering
\begin{tabular}{|c|c|l|}
\hline
$\kappa_3$ & cycle length & bifurcation type \\ \hline
1.0777760 & 98 & \multirow{2}{*}{{\bf DS(b)} $\quad A\to B$ } \\
1.0777761 & 98 & \\
\vdots & \vdots & \\
1.0777820 & 98 & \multirow{2}{*}{{\bf DS(b)} $\quad A\to B$ } \\
1.0777821 & 98 & \\
\vdots & \vdots & \\
1.0777905 & 98 & \multirow{2}{*}{{\bf DS(b)} $\quad A\to B$ } \\
1.0777906 & 98 & \\
\vdots & \vdots & \\
1.0777918 & 98 & \multirow{2}{*}{{\bf DS(b)} $\quad A\to B$ } \\
1.0777919 & 98 & \\
\vdots & \vdots & \\
1.0777935 & 98 & \multirow{2}{*}{{\bf DS(c)} $\quad A,b\to \emptyset$ } \\
1.0777936 & none & \\
\hline
\end{tabular}
\caption{Sequence of bifurcations in the bifurcation diagram for the $20$-dimensional model, as $\kappa_3$ is varied between $1.0777500$ and $1.0777936$.}\label{tab:bifseq}
\end{table}

\subsection{The bifurcation to chaos }
 \label{sec:bif_to_chaos}
Table~\ref{tab:bifseq} shows that there is a bifurcation of type ${\bf DS(c)}$, where the stable periodic orbit is lost, for some $\kappa_3$ in the interval $1.0777935< \kappa_3 <1.0777936$. Further inspection using the same algorithms as above shows that no bifurcation occurs over the first part of that interval, for $1.07779350<\kappa_3<1.07779359$.  Taking precisely the parameter value $\kappa_3=\kappa_4=1.07779359 $, 
we are on the last cycle with $98$ steps before the system becomes chaotic.
Call this cycle $A$.

The sequence of transitions is given by:
 17 ,    6,     1,    18,    19 ,    2 ,    7,     1,     2,     9,     8,     5,    10,    11 ,    6,     8,     7,    13 ,    5 ,    9,
 14,     6 ,   10,     7,    15,     9,    11,    17,    10 ,   16  ,  11,    13,    18,    14,    19,    15 ,   16,     1,  17,  13,
 2 ,   18 ,   14,    19,     3,    15,     1,    17,     5 ,   16 ,   20,     4 ,   17,     1,    13,     6,    14,     7,     4,    15,
 9 ,    1,    17,    16,    10,    13,     2,    18,    11,    14,     3 ,   19,    13 ,   14,    20,     1,     5,     2,    17,     6,
 3  ,  18 ,    7,     5 ,   19 ,    6,     9,     7,     9,     1,    20,     2,    17 ,    3,    18,    20 ,   19,     5.

This sequence is depicted graphically by the crosses in Figure~\ref{fig:96switchseq}.

As before, we calculate the return map, and compute the eigenvalues of the matrix in the map, 
which are the components of:
\begin{equation}
\alpha=\left(
{\tiny
\begin{array}{c}
168881223.00870542627357361792995 \\ 
167487916.0005497174842824483639 \\ 
-3452.2105762750412150403470619914 \\ 
327.2166778481279456146848973089 \\ 
22.13763477099014850083254767157 \\ 
3.966478490016213083317308464312 \\ 
1.0 \\ 
0.35271083665471966868291507985677 - 0.74240386041517346242259892950514 i \\ 
0.35271083665471966868291507985677 + 0.74240386041517346242259892950514i \\ 0.62356906224004248794632632719114 \\ 
-0.0560055763577170923036354826642 \\ 
0.021512395617491173358854892467027 \\ 
-0.0017231114534121761705722567792827 \\ 
0.0010224417047064596606404991739796 + 0.0011766950315849667533702998751534i\\ 
0.0010224417047064596606404991739796 - 0.0011766950315849667533702998751534i\\ 
-0.00025212369457840028515914363417561 + 0.00078720203285546231741676498196448i\\ 
-0.00025212369457840028515914363417561 - 0.00078720203285546231741676498196448i \\ 
0.00057723689603426681867258689004997 \\ 
-0.00042524279400878578158354560565191 \\ 
0 \\ 
\end{array}
}
\right)
\label{eq:lasteigenvalues}
\end{equation}
The fixed point for the cycle, a scaled version of the eigenvector, $V_\lambda$, that corresponds to the dominant eigenvalue
$\alpha_{max} = 168881223.0087$ 
is given by
\begin{equation}
v^*= (\alpha_{\max}-1)\frac{V_{\alpha}}{\varphi_{390}^{T} V_{\alpha}} 
{\tiny = 
\begin{pmatrix} 
   0.461314448510485\\
 -0.3572974061676364\\
  0.2882157063113082\\
  0.5094904987623157\\
                 0.0\\
 -0.4554529750157712\\
  0.5094854524724467\\
   0.510118602042323\\
 -0.4327293629423802\\
  0.5257436564792134\\
 -0.4970887614829581\\
  0.5777919811126316\\
  0.5556357272887729\\
 -0.4952560895668413\\
  0.5737054545732176\\
  0.5767092463885165\\
  0.1786939148402348\\
 -0.2171550075339816\\
0.009423760478470979\\
  0.1196197927784138
\end{pmatrix} }
\label{fixed55}
\end{equation}
where $v^*_5 = 0$ as expected. After moving the thresholds back to $\frac{1}{2}$, we get the fixed point in the original coordinates as
\begin{equation}
x^*=v^*+\frac{1}{2}=
{\tiny
\begin{pmatrix} 
   0.961314448510485\\
  0.1427025938323636\\
  0.7882157063113082\\
   1.009490498762316\\
                 0.5\\
 0.04454702498422882\\
   1.009485452472447\\
   1.010118602042323\\
 0.06727063705761984\\
   1.025743656479213\\
0.002911238517041899\\
   1.077791981112632\\
   1.055635727288773\\
0.004743910433158682\\
   1.073705454573218\\
   1.076709246388516\\
  0.6786939148402348\\
  0.2828449924660184\\
   0.509423760478471\\
  0.6196197927784138
\end{pmatrix} }
\label{eq:lastfixedpoint}
\end{equation}
The sequence of 247 alternative branching variables is (cycle step, alternative branching variable): 
$(1,  1)$, 
$(1,  6)$, 
$(2,  1)$, 
$(2, 18)$, 
$(3,  7)$, 
$(3, 18)$, 
$(4,  2)$, 
$(4,  7)$, 
$(5,  2)$, 
$(5,  7)$, 
$(6,  1)$, 
$(6,  7)$, 
$(7,  1)$, 
$(7,  3)$, 
$(8,  3)$, 
$(8,  8)$, 
$(8,  9)$, 
$(9,  3)$, 
$(9,  8)$, 
$(9,  9)$, 
$(10,  8)$, 
$(11, 10)$, 
$(12, 10)$, 
$(13,  6)$, 
$(14,  6)$, 
$(15,  8)$, 
$(15, 13)$, 
$(16,  7)$, 
$(16, 13)$, 
$(17,  5)$, 
$(17, 13)$, 
$(18,  5)$, 
$(18,  9)$, 
$(19,  9)$, 
$(19, 14)$, 
$(20,  6)$, 
$(20, 14)$, 
$(21,  6)$, 
$(21, 10)$, 
$(22, 10)$, 
$(22, 15)$, 
$(23,  7)$, 
$(23, 15)$, 
$(24, 11)$, 
$(24, 15)$, 
$(25,  9)$, 
$(25, 11)$, 
$(26, 11)$, 
$(26, 16)$, 
$(26, 17)$, 
$(27, 10)$, 
$(27, 16)$, 
$(27, 17)$, 
$(28,  8)$, 
$(28, 10)$, 
$(28, 12)$, 
$(28, 13)$, 
$(28, 16)$, 
$(29,  8)$, 
$(29, 12)$, 
$(29, 13)$, 
$(29, 16)$, 
$(29, 18)$, 
$(30,  8)$, 
$(30, 11)$, 
$(30, 12)$, 
$(30, 13)$, 
$(30, 18)$, 
$(31,  8)$, 
$(31, 12)$, 
$(31, 18)$, 
$(32, 18)$, 
$(33, 14)$, 
$(34, 19)$, 
$(35, 15)$, 
$(36,  1)$, 
$(36, 16)$, 
$(37,  1)$, 
$(37, 17)$, 
$(38, 13)$, 
$(38, 17)$, 
$(39,  2)$, 
$(39, 13)$, 
$(40,  2)$, 
$(40, 18)$, 
$(41, 14)$, 
$(41, 18)$, 
$(42,  3)$, 
$(42, 14)$, 
$(43,  3)$, 
$(43, 19)$, 
$(44,  3)$, 
$(44, 15)$, 
$(45,  1)$, 
$(45, 15)$, 
$(46,  1)$, 
$(46,  4)$, 
$(46,  5)$, 
$(46, 20)$, 
$(47,  4)$, 
$(47,  5)$, 
$(47, 16)$, 
$(47, 17)$, 
$(47, 20)$, 
$(48,  2)$, 
$(48,  4)$, 
$(48,  5)$, 
$(48, 16)$, 
$(48, 20)$, 
$(49,  2)$, 
$(49,  4)$, 
$(49, 16)$, 
$(49, 18)$, 
$(49, 20)$, 
$(50,  2)$, 
$(50,  4)$, 
$(50,  6)$, 
$(50, 18)$, 
$(50, 20)$, 
$(51,  2)$, 
$(51,  4)$, 
$(51,  6)$, 
$(51, 13)$, 
$(51, 18)$, 
$(52,  2)$, 
$(52,  6)$, 
$(52, 13)$, 
$(52, 17)$, 
$(52, 18)$, 
$(53,  1)$, 
$(53,  2)$, 
$(53,  6)$, 
$(53, 13)$, 
$(53, 18)$, 
$(54,  2)$, 
$(54,  6)$, 
$(54, 13)$, 
$(55,  6)$, 
$(56, 14)$, 
$(57,  7)$, 
$(58, 15)$, 
$(59,  9)$, 
$(59, 15)$, 
$(60,  1)$, 
$(60,  9)$, 
$(61,  1)$, 
$(61, 16)$, 
$(61, 17)$, 
$(62, 10)$, 
$(62, 16)$, 
$(62, 17)$, 
$(63,  2)$, 
$(63, 10)$, 
$(63, 16)$, 
$(64,  2)$, 
$(64, 10)$, 
$(64, 18)$, 
$(65,  2)$, 
$(65, 13)$, 
$(65, 18)$, 
$(66,  2)$, 
$(66, 11)$, 
$(66, 18)$, 
$(67, 11)$, 
$(67, 14)$, 
$(67, 18)$, 
$(68,  3)$, 
$(68, 11)$, 
$(68, 14)$, 
$(69,  3)$, 
$(69, 14)$, 
$(69, 19)$, 
$(70,  3)$, 
$(70, 13)$, 
$(70, 19)$, 
$(71, 13)$, 
$(71, 15)$, 
$(71, 19)$, 
$(72,  5)$, 
$(72, 13)$, 
$(72, 15)$, 
$(72, 20)$, 
$(73,  1)$, 
$(73,  5)$, 
$(73, 15)$, 
$(73, 20)$, 
$(74,  1)$, 
$(74,  5)$, 
$(74, 15)$, 
$(74, 20)$, 
$(75,  1)$, 
$(75,  5)$, 
$(76,  5)$, 
$(76, 17)$, 
$(77,  2)$, 
$(77, 17)$, 
$(78,  6)$, 
$(78, 17)$, 
$(79,  3)$, 
$(79,  6)$, 
$(80,  3)$, 
$(80, 18)$, 
$(81,  7)$, 
$(81, 18)$, 
$(82,  5)$, 
$(82,  7)$, 
$(83,  5)$, 
$(83, 19)$, 
$(84,  4)$, 
$(84,  8)$, 
$(84,  9)$, 
$(84, 19)$, 
$(85,  4)$, 
$(85,  6)$, 
$(85,  8)$, 
$(85,  9)$, 
$(86,  1)$, 
$(86,  4)$, 
$(86,  8)$, 
$(86,  9)$, 
$(86, 20)$, 
$(87,  1)$, 
$(87,  4)$, 
$(87,  7)$, 
$(87,  8)$, 
$(87, 20)$, 
$(88,  1)$, 
$(88,  4)$, 
$(88,  8)$, 
$(88, 10)$, 
$(88, 20)$, 
$(89,  1)$, 
$(89, 10)$, 
$(89, 20)$, 
$(90, 20)$, 
$(91,  2)$, 
$(92, 17)$, 
$(93,  3)$, 
$(94, 18)$, 
$(95,  5)$, 
$(95, 20)$, 
$(96,  5)$, 
$(96, 19)$, 
$(97,  5)$, 
$(97, 17)$, 
$(98,  1)$, 
$(98, 17)$. 

We notice that at every step there exists at least one alternative branching variable. Some of the returning cone conditions evaluated at the fixed point in Equation~\eqref{fixed55} are given in Table~\ref{TC4}.
The full table has $247$ conditions, one for each alternative branching variable around the cycle. The second column in the table represents the alternative branching variables at each step.

\begin{table}[h]
\begin{equation*}
\begin{array}{|c|c|l|r|}
\hline
Steps  &&\texttt{The returning Cone Conditions at the fixed point in~\eqref {fixed55} }
\\
\hline
1& 
 \begin{aligned} 
 1\\
 6 \end{aligned}&
 \begin{aligned} 
 &  0.56524106734050055890839226148344 >0,\\
 &  0.45592105427626472120467532320625 >0.
 \end{aligned}\\
\hline
2&
 \begin{aligned}    
 1\\
 18\end{aligned}&                                                        
  \begin{aligned}
 & 0.10932001306423583770371693827719 >0,\\
 & 0.25095202139197877720038109139305 >0.
 \end{aligned}\\
\hline
3& 
 \begin{aligned}  
 7  \\
 18\end{aligned}&                                           
 \begin{aligned}
 & 1.8495002115611672789677868293544 >0,\\
 & 0.14163200832774293949666415311585 >0.
 \end{aligned}\\
\hline
4&
 \begin{aligned}    
 2\\
 7\end{aligned} & 
 \begin{aligned}
 &  1.3201748750260453280724218855438 >0,\\
 &  1.7078682032334243394711226762386 >0.
 \end{aligned}\\
\hline
5&
 \begin{aligned}
 2\\
 7\end{aligned} & 
 \begin{aligned}
 & 0.071481095672937177506166839729021 >0,\\
 & 0.45917442388031618890486763042381  >0.
 \end{aligned}\\
\hline
6&
 \begin{aligned}
 1\\
 7\end{aligned}& 
 \begin{aligned}    
 &  1.1698812147563941529071538736019 >0,\\
 &  0.38769332820737901139870079069478 >0.
 \end{aligned}\\
\hline
7&\begin{aligned}
 1\\
 3\end{aligned}& 
 \begin{aligned}    
 &  0.78218788654901514150845308290711 >0,\\
 &  2.9441615037496036734924965683991 >0.
 \end{aligned}\\
\hline
8& 
 \begin{aligned}
 3\\
 8\\
 9 \end{aligned}&    
 \begin{aligned}    
 &  2.161973617200588531984043485492 >0,\\
 &  3.4414861767179768129615448058359 >0,\\
 &  2.4656583942250006375763220726624 >0.
 \end{aligned}\\
\hline
9& \begin{aligned}
 3\\
 8\\
 9\end{aligned} & 
 \begin{aligned}    
 & 0.85170665667342708072803114705784 >0,\\
 & 2.1312192161908153617055324674018 >0, \\
 & 1.1553914336978391863203097342283 >0.
 \end{aligned}\\
 \hline	... & &...... \\
 \hline	 ... & &......  \\
\hline	
88&   
 \begin{aligned} 
 1\\ 4\\8\\10\\20\end{aligned}&     
 \begin{aligned}
 &  4581660.6769453748233314540617678 >0,\\
 &  1704.4472094592544609482497293831 >0\\
 &  213854.14741630652719042714153496 >0,\\
 &  16633521.50645180283211336549775  >0,\\
 &  7728774.9991587775195405320264413 >0.
 \end{aligned}
\\
\hline		... & &...... \\
\hline	 . .. & &......  \\
\hline 
98& 
 \begin{aligned} 1\\ 17\end{aligned}&     
 \begin{aligned}
 & 155814696.512074361391629424965 >0,\\
 & 60356093.764864605132467806668482>0.
 \end{aligned}\\
\hline
\end{array}
\end{equation*}
\caption{Examples taken from the 247 inequalities satisfied by cycle A's fixed point.}\label{TC4}
\end{table}

All are positive, so the fixed point is in the returning cone and we have a periodic orbit.

Now we can vary $\kappa_3$ ($=\kappa_4$) until at least one condition for the fixed point to lie in the returning cone is violated, {\em i.e.}, at least one component $R_{i}v^*<0$, $i \in \{1,..., k\}$, where $k$ is the number of conditions. Note that both $R$ and $v^*$ are functions of $\kappa_3$. At the value of $\kappa_3$ where a component $R_{i}v^*=0$, the cycle (with this fixed point) is lost, and we have a {\bf DS} bifurcation. 

\begin{figure}%
\includegraphics[width=.8\columnwidth]{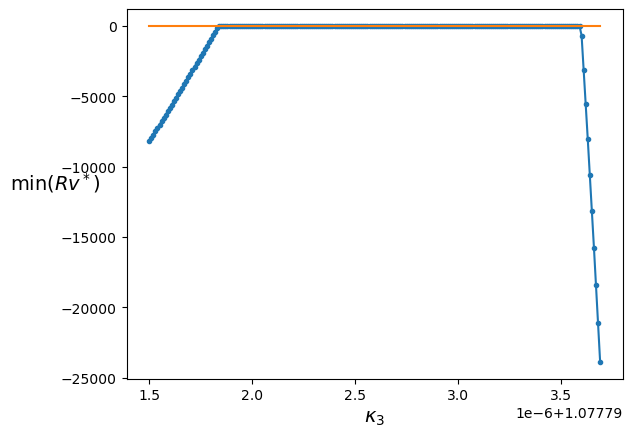}%
\caption{The minimum of the returning cone $\min\{ R v^*\}$ at each fixed point as a function of $\kappa_3$ ($=\kappa_4$). The value of the minimum is $>0$ but small for $1.07779184 \le \kappa_3 \le 1.07779359$.} %
\label{figure23}%
\end{figure}

\begin{figure}%
\includegraphics[width=.8\columnwidth]{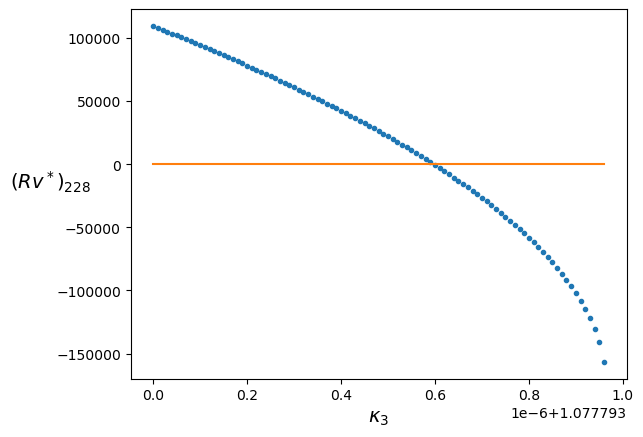}%
\caption{The $228^{th}$ component of the returning cone condition, $Rv^*$, that corresponds to cycle step $88$ with the alternate switching variable $4$ as a function of $\kappa_3$. The right-hand part of this curve corresponds to the right-hand branch of the curve in Figure~\ref{figure23}. At lower values of $\kappa_3$ a different condition is minimal.} %
\label{figure22}%
\end{figure}

Plots of the minimum returning cone condition, $\min\{ R v^*\}$, at the fixed point, as a function of $\kappa_3$, are given in Figures~\ref{figure23} and \ref{figure22}. The cycle exists in the interval $\left[ 1.07779184, 1.07779359\right]$. To the left of the left boundary of this interval we detect another cycle also with length $98$. To the right of the right boundary of this interval, simulations show that no stable periodic orbit exists. The trajectory from the fixed point hits a double switching at the $88^{th}$ step of the cycle, where variable 4 ($u_1$) switches at the same time as variable 7 ($z_2$). The latter, which occurs before the bifurcation, keeps us on cycle $A$; the former, which occurs after the bifurcation, means we must leave the cycle, and no periodic orbit now exists for cycle $A$. Thus after $\kappa_3 = 1.07779359$ the cycle is lost and the system must either follow a different cycle or become chaotic.

%
%

\subsection{A proof that no stable periodic orbit exists (locally) after the bifurcation}
\begin{figure}
\centering
  \includegraphics[width=.7\columnwidth]{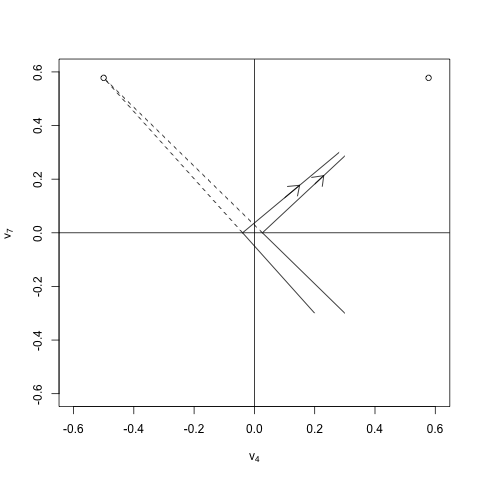}%
\caption{Sketch of the phase plane in the two variables $v_4$ and $v_7$, at step 88 of the cycle, where they switch simultaneously at the bifurcation point. Before the bifurcation, only $v_7$ switches; after the bifurcation $v_4$ switches first, then $v_7$, then $v_4$ switches back again. The circles are the focal points.}%
\label{fig:phaseplot47}%
\end{figure}

Increasing  $\kappa_3$ (and $\kappa_4=\kappa_3$), the $228^{th}$ component of the returning cone condition (ordered in the obvious way by cycle steps first and variable number within each step), which corresponds to cycle step $88$ with alternate switching variable $4$, becomes negative $-686.471537$ around the value $\kappa_3=\kappa_4=1.0777936$. At step $88$,
alternate switching variable $4$ is taken when the cycle above is lost, while variable 7 switched at step $88$ on the cycle before the bifurcation point. At the $88^{th}$ step, we detect that the trajectory is in the box  $(0     1     0     1     1     0     0     1     1     1     0     1     1     0     1     1     0  1     0     1)$. Now examining the phase plane in those two variables, $4$ and $7$, the focal points for the four boxes around the threshold intersection can be determined (as functions of parameters) to be as shown in Table~\ref{tab:phplane}. 
\begin{table}[h]
\begin{equation*}
\begin{array}{|c|c|c|}
 \hline
Variables &  4 & 7 \\ 
 \hline
10 &-\theta&   \frac{k_3}{\gamma}- \theta      \\
\hline
00 &-\theta &    \frac{k_4}{\gamma}- \theta  \\ 
 \hline
01 &\frac{k_4}{\gamma}- \theta & \frac{k_3}{\gamma}- \theta\\ 
\hline
11 & \frac{k_4}{\gamma}- \theta & \frac{k_3}{\gamma}- \theta\\ 
 \hline
\end{array}
\label{T4boxes}
\end{equation*}
\caption{Focal point coordinates corresponding to Figure~\ref{fig:phaseplot47}.} \label{tab:phplane}\end{table}
The corresponding phase portrait, with $\kappa_3=\kappa_4=1.0777936$, is shown in Figure~\ref{fig:phaseplot47}.

Starting from the $88^{th}$ step, before the bifurcation we then follow the switching sequence  7 $\longrightarrow$   9$\longrightarrow$     1$\longrightarrow$    20 $\longrightarrow$     2 $\longrightarrow$    17 $\longrightarrow$    3$\longrightarrow$    18$\longrightarrow$    20 $\longrightarrow$   19$\longrightarrow$     5 $\cdots$. After the bifurcation, the system follows the sequence  4 $\longrightarrow$ 7 $\longrightarrow$4 $\longrightarrow$    9$\longrightarrow$     1$\longrightarrow$    20$\longrightarrow$     2$\longrightarrow$    17 $\longrightarrow$    3$\longrightarrow$    18$\longrightarrow$    20 $\longrightarrow$   19$\longrightarrow$     5 $\cdots$. Now, the $4^{th}$ variable switches first, then the $7^{th}$, then the $4^{th}$ switches back again and the course of the original cycle is picked up again, giving a cycle of length $100$. Call this cycle $B$.

It is clear from the phase plane analysis above that we have an unambiguous double-switching bifurcation. To determine the nature of the cycles on either side of the bifurcation, we need to apply the results of Section~\ref{sec:bifurcations}, using the eigenvalues of the matrices of cycles $A$ and $B$ evaluated at the bifurcation point. Of course, it is not possible to compute the bifurcation point with infinite precision, so we look at nearby parameter values on either side of the bifurcation.

Taking the original cycle ($A$) at $\kappa_3=1.07779359$, just before the bifurcation point, the eigenvalues of the matrix in the cycle map are the components of~\eqref{eq:lasteigenvalues},
the eigenvector that corresponds to the dominant eigenvalue 
$$\tiny \alpha_{max} =   168881223.008735\,,$$
and the returning cone conditions are all satisfied, as shown above, so the cycle exists and is stable. 

Taking the original cycle ($A$) at a parameter value, $\kappa_3=1.0777936$, very close to (but just after) the bifurcation point, we calculate the map and the eigenvalues of the matrix in the map, given by components of
\begin{equation*}
\tiny
\alpha=\left(
\begin{array}{c}
 168871811.20689127045934164009889\\ 
 167497245.41252679624479226891464\\ 
 -3452.2106265890646335589461971197\\ 
 327.21669909120019455101506529124\\ 
 22.137635218231384359990709281584\\ 
 3.966478451724300657393367357525\\ 
 1.0 \\ 
 .35271084934237772713247712826879 + 0.74240383278823543289873495825401i\\ 
 .35271084934237772713247712826879 - 0.74240383278823543289873495825401i\\ 
 0.62356906642878674480051470905631 \\ 
-0.056005576584881301419538965924618\\ 
 0.021512395817243413126229827093559\\ 
-0.0017231114521899393280387190179325\\ 
 0.0010224417062848763520996372872419 + 0.0011766950218005976175947497700139i \\ 
0.0010224417062848763520996372872419 - 0.0011766950218005976175947497700139i\\ 
-0.00025212369755419588036724905238397 + 0.00078720202887089098723687292995249i\\ 
-0.00025212369755419588036724905238397 - 0.00078720202887089098723687292995249i\\ 
0.00057723688917951025076212166094329\\ 
-0.00042524279251787835155905918099087\\ 
0.0
\end{array}
\right)
\end{equation*}
and, the eigenvector that corresponds to the dominant eigenvalue 
$$\tiny \alpha_{max}=\alpha_1 = 168871811.20689 $$
is given, once normalized to be a fixed point, by
\begin{equation*}
v^*= (\alpha_{\max}-1)\frac{V_{\alpha}}{\varphi_{A}^{T} V_{\alpha}}=
\tiny\left(
\begin{array}{c}
 0.46132968382339546002648578512549\\
-0.35729603610243751377443305062702\\
  0.2882141755743414445564467454462\\
 0.50947347803158150187011136748411\\
                                0.0\\
-0.45544802406504469766376038045597\\
 0.50947551055422150427095404760315\\
 0.51010160425877461972551729340223\\
-0.43271261052788895757485785405955\\
 0.52571093089168391671243529327864\\
-0.49708865532607725842287491773131\\
 0.57779199111283439544720155118459\\
 0.55563559443514800576816016094903\\
-0.49525601762715111009210791341817\\
 0.57370562294702872250291451391012\\
 0.57670920116858155659074689540483\\
 0.17869401374348360448292731120934\\
 -0.2171505288377642403853602765684\\
0.0094148345185094996108480066164826\\
 0.11961830795789301786374847204782
\end{array}
\right)\,.
\end{equation*}
Both eigenvalues and dominant eigenvector are slight perturbations of the pre-bifurcation values. It is not necessary to pinpoint their values exactly {\em at} the bifurcation value; it is clear that the eigenvector corresponding to the periodic orbit is still the dominant one there, {\em i.e.}, $\sigma_\alpha^+=\sigma_\alpha^-=0$. We will always have $\sigma_{\alpha}^+ = \sigma_{\alpha}^- = 0$ when we start with a stable cycle $A$ before the bifurcation.

The new cycle ($B$) that exists after the bifurcation point has two additional switches of the $4^{th}$ variable (as shown by the phase plane corresponding to Table~\ref{T4boxes}, in Figure~\ref{fig:phaseplot47}), and so is of length $100$. At a parameter value ($\kappa_3=1.07779359$) just before the bifurcation value, the eigenvalues of the matrix in the map of this new cycle are the components of
\begin{equation*}
\tiny
\beta=\left(
\begin{array}{c}
263342326.12720817539010376250291\\ 168875233.52460524403567401433559\\ 39855.344897114382982395216954362\\ -391.89723819192094947624083796032\\ 7.886312534774945079338576647601\\ 5.3971433581975478076546983129435\\ 1.0\\ (0.28819310545530858655996603188492 - 0.53848729234595482714447712166693i)\\ (0.28819310545530858655996603188492 + 0.53848729234595482714447712166693i)\\ 0.23591450153846613298869600261073\\ -0.050292729365532798124271814987318\\ 0.011437616316420404128283570518223\\ -0.0017276370984439396118086025211955\\ (0.00099552541085193402184772601603794 + 0.00090181937604378308942072253694127i)\\ (0.00099552541085193402184772601603794 - 0.00090181937604378308942072253694127i)\\ (-0.00023296136585379161243132935995653 - 0.00085358033860887696425524766685551i)\\ (-0.00023296136585379161243132935995653 + 0.00085358033860887696425524766685551i)\\ 0.00060513636557464959011265170358774\\ -0.00045244784150126532773117448983143\\
0.0
\end{array}
\right)\,.
\end{equation*}
The second-largest eigenvalue is 
$$\tiny \beta_2 =  168875233.524605  $$
and the eigenvector corresponding to this eigenvalue, once normalized to be a fixed point, is
\begin{equation*}
v^*= (\beta_2-1)\frac{V_{\beta}}{\varphi_{B}^{T} V_{\beta}}=
\tiny\left(
\begin{array}{c}
  0.46132094740474049439757091227888\\
 -0.35729648757276521924989449613817\\
  0.28821467056085603189184567080638\\
  0.50947848198849590928440545373797\\
                                 0.0\\
 -0.45544959034402810072092280237422\\
  0.50947855607847699620909828500889\\
  0.51010659907115565751550723114893\\
  -0.4327175424689104242775339380044\\
  0.52572013764270614204907043658246\\
 -0.49708869765759250734598308864904\\
  0.57779198111996068893464512965684\\
  0.55563565068302559835996286208514\\
 -0.49525605888895482669052958027207\\
  0.57370558809214491974256891962863\\
  0.57670921125158893898112250234312\\
  0.17869510780208721229353666440284\\
 -0.21715301328177869014603137545475\\
0.0094198036976830061447779643541338\\
  0.11961849434983196368108081925823
\end{array}
\right)\,,
\end{equation*}
which is a small perturbation of the eigenvector on which the fixed point of the cycle map lies, and these must coincide exactly right at the bifurcation value. Since this is no longer the dominant eigenvector, the corresponding periodic orbit is unstable if it exists.

Now, consider this $100$-step cycle ($B$) at a parameter value  ($\kappa_3=1.0777936$) after the bifurcation point. 
Calculating the matrix of the cycle map, we compute its eigenvalues as components of 
\begin{equation*}
\beta=
\tiny\left(
\begin{array}{c}
263343885.74081094753100674251093\\
168874225.4802148397490450842695\\
39855.346564911425267560100845684\\
-391.89724316541658174538029374513\\
7.8863145967290459238841784133889\\
5.3971421203660983336686075820863\%1.0\\
1.0\\
0.28819311540899026215632030612813 + 0.53848728366186300227939989673818i\\
0.28819311540899026215632030612813 - 0.53848728366186300227939989673818i\\
0.23591450363977853388464429192625\\
-0.050292728880813911075948999272475\\
0.011437616363551421502638582273448\\
-0.0017276370969445077739647307541431\\
0.00099552541293140142446486000940216 - 0.00090181936594346091124464441687514i\\
0.00099552541293140142446486000940216 + 0.00090181936594346091124464441687514i\\
-0.00023296136924317965809093098523186 + 0.00085358033374864300558694866549666i\\
-0.00023296136924317965809093098523186 - 0.00085358033374864300558694866549666i\\
0.00060513635756515239603226242351698\\
-0.00045244784012686662357192919326784\\
0                 
\end{array}
\right)\,,
\end{equation*}
again a small perturbation of the pre-bifurcation value. Again, the eigenvalue corresponding to the periodic orbit (now lost) is the second one. The second eigenvector (normalized) is
\begin{equation*}
v^*= (\beta_2-1)\frac{V_{\beta}}{\varphi_{B}^{T} V_{\beta}}=
\tiny\left(
\begin{array}{c}
  0.46132705434569942046510982542177\\
 -0.35729640624584849791391854595816\\
  0.28821459290960245794798610702477\\
  0.50947831842708983892262734189592\\
                                 0.0\\
 -0.45544938772062072808938154657683\\
  0.50947828878779777436919939901158\\
  0.51010643908799481344741352321517\\
 -0.43271737185392540033625096492224\\
  0.52572040318159843909985547786395\\
 -0.49708868106544165563237852460043\\
  0.57779199110990174388025285390911\\
  0.55563562533898306918266535720004\\
 -0.49525603003893662687622728272342\\
    0.573705569223721811273105514724\\
  0.57670921533232002594990620027634\\
  0.17869353629328043943627372020129\\
 -0.21715133533454720149626059232566\\
0.0094164347674689914014505044531332\\
  0.11961883028911383994503897022451
  \end{array}\right)\,,
\end{equation*}
again a small perturbation of the pre-bifurcation eigenvector.
Thus, at the bifurcation point we have $\beta_2\approx 168875233.524605$ and $\sigma_\beta^+=1$ (since $\beta_1>\beta_2$), while $\sigma_\beta^-=0$ (there are no real eigenvalues less than $-\beta_2$). 

Then,
$\sigma_\alpha^- + \sigma_\beta^- =0$ is even, which means that no period doubling occurs. But $\sigma_\alpha^+ + \sigma_\beta^+ = 1$ is odd, which means that merging and disappearance of two orbits occurs. The bifurcation is of type {\bf DS(c)} and has the behaviour $A,b\to\emptyset$.

Thus, by Proposition~\ref{prop:DSbif} the two periodic orbits exist before the bifurcation, but at the bifurcation point they collide and after the bifurcation point both disappear (like a saddle-node bifurcation of cycles). 

These results can be verified manually. We already know that the periodic orbit for cycle $A$ exists and is stable before the bifurcation.
None of the eigenvalues of $A$ after the bifurcation point that are greater than 1 have eigenvectors that satisfy the returning cone conditions for existence of the cycle, so this periodic orbit is lost. 
 
The periodic orbit of the $B$ cycle exists before the bifurcation if the eigenvector lies in the starting boundary and if the returning cone conditions are all positive.
The sequence of alternative branching variables is similar to that of cycle $A$ until the $228^{th}$ condition. Starting from the $228^{th}$, the sequence of alternative branching variables is:
$(  88,  7)$,
$(  88,  8)$,
$(  88, 10)$,
$(  88, 20)$,
$(  89,  8)$,
$(  89, 10)$,
$(  89, 20)$,
$(  90,  9)$,
$(  90, 10)$,
$(  90, 20)$,
$(  91,  1)$,
$(  91, 10)$,
$(  91, 20)$,
$(  92, 20)$,
$(  93,  2)$,
$(  94, 17)$,
$(  95,  3)$,
$(  96, 18)$,
$(  97,  5)$,
$(  97, 20)$,
$(  98,  5)$,
$(  98, 19)$,
$(  99,  5)$,
$(  99, 17)$,
$( 100,  1)$,
$( 100, 17)$.

The returning cone conditions evaluated at this fixed point have all components positive, so the cycle exists, but the eigenvalue is not the dominant one, so it is unstable.

After the bifurcation value, none of the eigenvalues of the $B$ matrix greater than 1 have eigenvectors satisfying the returning cone conditions, so there is no periodic orbit after the bifurcation point. 

Thus, two periodic orbits exist before the bifurcation, one stable, one unstable, but at the bifurcation point they collide and after the bifurcation point both disappear. 
We cannot say anything definitive about the flow after the bifurcation point, except that locally, periodic behaviour is lost.

\section{Discussion}\label{sec:discussion}
 The main message we wish to convey in this paper is that Glass networks are a class of dynamical systems in which it is possible to prove existence, stability, and bifurcations of periodic orbits in high dimensions. This is remarkable. Computer aid is needed to compute the maps and conditions that need to be checked, though these can in principle be done by hand (and can actually be done for low-dimensional examples). Computer aid is definitely needed to compute eigenvalues and eigenvectors of the matrices involved, but this is just to confirm conditions for the theorems guaranteeing existence and stability of periodic orbits; essentially, the approach is rigorous.
 
 Even when we integrate a Glass network `numerically', such as in computation of the bifurcation diagrams, we are actually calculating explicit wall-to-wall transition maps, not using an approximate numerical integration scheme (like a Runge-Kutta method). 
 
 We illustrate the methods for analysis of Glass networks here in an example with an important application in cybersecurity: designs for TRNGs that are robust because intrinsically chaotic. The existence of parameter intervals on which the behaviour was `chaotic' by the criterion of numerically estimated Lyapunov exponents was suggestive, but here we show explicitly that a transition from a periodic window of parameter values to a chaotic one occurs where a periodic orbit crosses the boundary of its returing cone, and thus ceases to exist, in a kind of non-smooth version of a saddle-node bifurcation of cycles: a stable and unstable cycle collide and annihilate leaving no periodic orbit, at least locally in phase space, on the other side of the bifurcation. This occurs when our bifurcation parameter $\kappa_3\approx 1.0777936$.

 The existence of this type of bifurcation in itself is not enough to guarantee that chaos ensues. In fact, another bifurcation of this type occurs at an earlier parameter value ($\kappa_3\approx 1.0777603$), but in that instance, there is another stable cycle elsewhere in phase space for the trajectory to fall onto, after the bifurcation. However, after the final value ($\kappa_3\approx 1.0777936$), there do not appear to be any more stable cycles and the system appears to become chaotic.

While we have shown that a stable cycle is lost at the apparent transition to chaos, we have not shown directly that the dynamics after the bifurcation is chaotic, or even aperiodic. There are methods for showing that in a given network there is an attractor on which the dynamics must be aperiodic, though these have only been applied in four dimensions~\cite{edwards2001chaos,edwards2005matrices}. They would be more difficult to apply in $20$ dimensions. Chaos has also been proven to exist rigorously, but only in an example of dimension $3$ with unequal decay rates~\cite{li2006chaotic}, and in an example of dimension $3$ with multiple thresholds per variable (or equivalently, of dimension $6$ with single thresholds per variable)~\cite{edwards2012explicit}. The methods for these specific examples depend on horseshoe-like structures and are not easily generalizable.

\begin{figure}
\centering
  \includegraphics[width=\columnwidth]{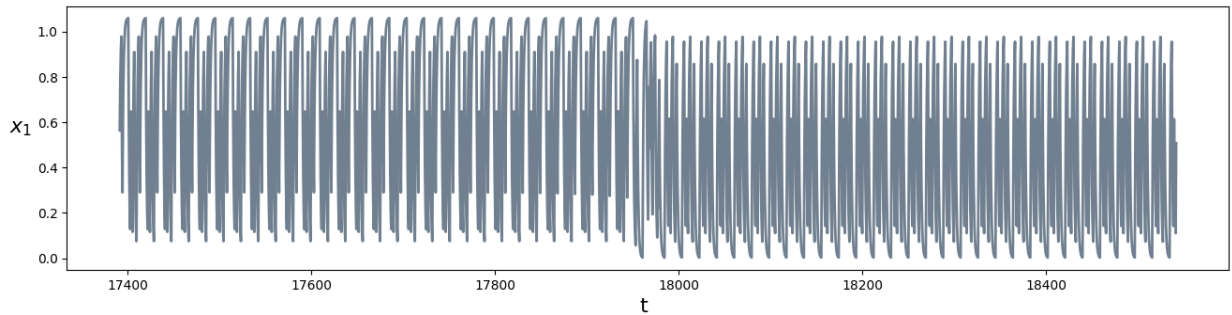}%
\caption{Time series of variable $x_1$ at the parameter value $\kappa_3=1.07779360$, immediately after the final bifurcation point. The trajectory clearly lies close to a periodic orbit and repeats its cycle many times before exiting the returning cone and soon falling onto another almost-periodic cycle.}%
\label{fig:intermittency}%
\end{figure}

 While the possibility of rigorous analysis in high dimensions does make this class of systems remarkable, we cannot claim that it is always easy. One of the issues that arises in tracking the bifurcations, is that immediately after a double-switching bifurcation, the (previously) stable fixed point of a cycle map has fallen just outside its returning cone, and so is now only a stable pseudo-fixed point, but it is still very close to the boundary. It no longer exists as a fixed point, but if there is no periodic orbit after the bifurcation (such as in the $A,b \to \emptyset$ type), then the pseudo-fixed point continues to attract nearby trajectories. Thus, nearby trajectories {\em inside} the returning cone can follow the cycle a very large number of times, gradually converging towards the pseudo-fixed point, before eventually leaving the returning cone. See Figure~\ref{fig:intermittency}. This phenomenon may contribute to the dense regions within the chaotic bands of the bifurcation diagrams (Figs.~\ref{fig:bif_103_1095_x1}, \ref{fig:bif_107775_107781}). It can also cause computer code to falsely conclude that a periodic sequence of boxes has been found numerically. However, the tools outlined in this paper allow us in such cases to determine that no periodic orbit exists through such a cycle of boxes.
 
Our example of the ring circuit proposed by Scott Best has some special problems because of the symmetries of the system. This leads to cycles in which the different units are doing almost the same thing at different phases of the cycle, though not necessarily exactly the same thing (for example, in a cycle of 96 switching steps, not all of the $5$ units can be doing a phase-shifted version of the same thing at all phases of the cycle, or the number of steps would be a multiple of $5$), so there may be a set of bifurcations that occur at almost exactly, or even exactly, the same parameter value. For example, this phenomenon seems to occur at $\kappa_3$ a little above $1.02118$, where as we {\em decrease} the parameter, a period doubling bifurcation occurs (type {\bf DS(d)}, {\bf DS(e)} or {\bf DS(f)}), but tracking it is difficult because four other bifurcations occur at the same or almost the same parameter value. To the right of the bifurcation, we have a stable cycle of length $210$. To the left of the bifurcation, there is a stable cycle of length $430$. Note that the period can double even if the number of switchings does not exactly double, and there are several bifurcations occurring together here. The period doubling is evident in Figure~\ref{fig:bif_100_105}, though the fine details cannot, of course, be seen.

We used $64$-decimal digit precision to handle the large numbers that appear in the calculations, and differences of many orders of magnitude. This was probably overly cautious, and somewhat lower precision could have been used. On the other hand, it suggests that going beyond $20$ dimensions and cycles of length $400$ should also be feasible. Furthermore, in principle extended precision calculations can be done with arbitrarily many digits, though computation times increase. All computations for the current work were done on a laptop computer. The most time-consuming calculations were the generation of the bifurcation diagrams, which took up to a few hours to compute. Similar or longer computation times would be expected for any $20$-dimensional nonlinear system of ODEs. Bifurcation diagrams are essentially based on numerical simulations, although we are here computing wall-to-wall maps explicitly. However, it is worth pointing out that the tracking of periodic orbits and identifying their bifurcation types, as done in Section~\ref{sec:bif_to_chaos}, is analytic (not based on simulations) and much faster to compute: just a few minutes for the parameter ranges done here. Thus, in principle, bifurcations could be tracked in systems of very high dimension, such as might occur in deep-learning networks, by the methods described here.

\section*{Appendix A}
For the $2n$-unit network~\eqref{eq:2n2v2} with $n=5$ and symmetric parameters (identical units) considered in Section~\ref{sec:10d}, the map for the first step of the cycle was given by Equations~\eqref{eq:firststep}. The second step is given by
\begin{equation}
B^{(2)}=
\begin{pmatrix} 
 1& 0& 0& 0& 0& 0&          -\frac{1}{d_1} & 0& 0& 0\\
 0& 1& 0& 0& 0& 0& -\frac{d_2}{d_1}& 0& 0& 0\\
 0& 0& 1& 0& 0& 0&                       -1& 0& 0& 0\\
 0& 0& 0& 1& 0& 0& -\frac{d_2}{d_1}& 0& 0& 0\\
 0& 0& 0& 0& 1& 0&         -\frac{1} {d_1}& 0& 0& 0\\
 0& 0& 0& 0& 0& 1&          -\frac{1} {d_1}& 0& 0& 0\\
 0& 0& 0& 0& 0& 0&                        0& 0& 0& 0\\
 0& 0& 0& 0& 0& 0& -\frac{d_2}{d_1}& 1& 0& 0\\
 0& 0& 0& 0& 0& 0&                       -1& 0& 1& 0\\
 0& 0& 0& 0& 0& 0& -\frac{d_2}{d_1}& 0& 0& 1
\end{pmatrix}, \quad \psi^{(2)}=
\begin{pmatrix} 
0\\ 0 \\ 0\\ 0\\ 0\\ 0\\ \frac{2}{d_1} \\ 0\\ 0\\ 0
\end{pmatrix} ,
\end{equation}
where $d_1=1-2\kappa_1$ and $d_2=1-2\kappa_2$.

If the result of these two steps leads us to a permutation of the initial point corresponding to a rotation two units around the cycle, then, because of the 5-fold symmetry, continuing for 8 further steps, we must return to the initial point, which is then a fixed point for the cycle map. On the other hand, if after the first two steps we rotate back two units, we will recover the initial point immediately. 
Let $P$ be the matrix of the rotation $\left(\begin{array}{c}
1,2,3,4,5,6,7,8,9,10\\
7,8,9,10,1,2,3,4,5,6
\end{array}\right)$ in $\R^{10}$ as follows:
\begin{equation}
P=
\begin{pmatrix} 
 0&    0& 0& 0& 1& 0& 0& 0& 0& 0 \\
 0& 0& 0& 0& 0& 1& 0& 0& 0&  0\\
 0& 0& 0& 0& 0& 0& 1& 0& 0&  0\\
 0& 0& 0& 0& 0& 0& 0& 1& 0&  0\\
 0& 0& 0& 0& 0& 0& 0& 0& 1&  0\\
 0& 0& 0& 0& 0& 0& 0& 0& 0& 1    \\
 1& 0& 0& 0& 0& 0& 0& 0& 0&  0\\
 0& 1& 0& 0& 0& 0& 0& 0& 0&  0\\
 0& 0& 1& 0& 0& 0& 0& 0& 0&  0\\
 0& 0& 0& 1& 0& 0& 0& 0& 0& 0
\end{pmatrix} .
\end{equation}
Then, the composition of the mappings for the first two steps with this rotation gives the matrix (with third row and column removed, since $v_3=0$ on the starting wall):
\begin{equation}
B_P=\left. PB^{(2)}B^{(1)}\right|_{(3)}=
\begin{pmatrix} 
0& 0&  0&                     \frac{1}{d_1}& 0&          -\frac{1}{d_1}& 0& 0& 0\\
0& 0&  0&    \frac{1}{d_1} - d_2& 1&          -\frac{1}{d_1}& 0& 0& 0\\
0& 0&  0& \frac{2\kappa_1 d_2}{d_1}& 0& -\frac{d_2}{d_1}& 1& 0& 0\\
0& 0&  0&                           2\kappa_1& 0&                       -1& 0& 1& 0\\
0& 0&  0& \frac{2\kappa_1 d_2}{d_1}& 0& -\frac{d_2}{d_1}& 0& 0& 1\\
1& 0&  0&                  \frac{2\kappa_1}{d_1}& 0&          -\frac{1}{d_1}& 0& 0& 0\\
0& 1&  0& \frac{2\kappa_1 d_2}{d_1}& 0& -\frac{d_2}{d_1}& 0& 0& 0\\
0& 0&  0&                           2\kappa_1& 0&                       -1& 0& 0& 0\\
0& 0&  1& \frac{2\kappa_1 d_2}{d_1}& 0& -\frac{d_2}{d_1}& 0& 0& 0
\end{pmatrix} .
\end{equation}
It can be shown that the eigenvalues of $B_P$ do not depend on $\kappa_2$. The proof is in Appendix B.
The fixed point and the returning cone, however, do depend on $\kappa_2$.

Taking the same parameter values as in Equation~\eqref{eq:kappas} in Section~\ref{sec:10d}, $\kappa_1=1.53, \kappa_2=2$, we get
\begin{equation}
B_P=
\begin{pmatrix} 
0&    0&  0&-0.4854&  0&0.4854&  0&  0&  0\\
0&    0&  0&  2.515&1.0&0.4854&  0&  0&  0\\
0&    0&  0&  4.456&  0&-1.456&1.0&  0&  0\\
0&    0&  0&   3.060&  0&  -1.0&  0&1.0&  0\\
0&    0&  0&  4.456&  0&-1.456&  0&  0&1.0\\
1.0&  0&  0& -1.485&  0&0.4854&  0&  0&  0\\
0&    1.0&    0&  4.456&  0&-1.456&  0&  0&  0\\
0&    0&  0&   3.060&  0&  -1.0&  0&  0&  0\\
0&    0&1.0&  4.456&  0&-1.456&  0&  0&  0
\end{pmatrix},\quad
\psi=
\begin{pmatrix} 
 0\\0\\ 0\\ 2.971\\ 0\\ -0.9709\\ 0\\ 0\\ 0
\end{pmatrix}.
\end{equation}
The eigenvalues of this matrix $B_P$ are the components of
\begin{equation*}
\lambda=
\begin{pmatrix} 
               4.303\\
                 1.0\\
              0.5462\\
     0.309 + 0.9511i\\
     0.309 - 0.9511i\\
 - 0.6518 + 0.02664i\\
 - 0.6518 - 0.02664i\\
   - 0.809 + 0.5878i\\
   - 0.809 - 0.5878i                                                   
\end{pmatrix},
\end{equation*}
\begin{equation*}
\end{equation*}
and thus, the dominant eigenvalue is $\lambda_{\max}=4.303$, the corresponding eigenvector is 
\begin{equation*}
w_{\max}=
\begin{pmatrix} 
 -0.04747\\
  0.26254\\
  0.44230\\
  0.29397\\
  0.45201\\
 -0.12682\\
  0.40840\\
  0.23854\\
  0.45018
\end{pmatrix} ,
\end{equation*}
and the fixed point for this cycle map is given by
\begin{equation}
v^*= \frac{(\lambda_{\max}-1)}{\psi^{T} w_{\max}} w_{\max} = 
\begin{pmatrix} 
-0.157351\\
   0.87017\\
   1.46598\\
  0.974368\\
   1.49816\\
 -0.420367\\
   1.35362\\
  0.790624\\
   1.49209
\end{pmatrix} ,
\label{fixed2}
\end{equation}
where $v_3^*=0$, the same fixed point as in Equation~\eqref{eq:fixedpoint10}. This map is of course different than the map for the full 10-step cycle, but a fixed point of this map must be a fixed point of the full cycle map because of the symmetry of the circuit. Also, the returning cone for this map has just one condition (inequality) because only one alternate exit variable occurs in the first two steps of the cycle. Since a fixed point for this map is a fixed point for the full cycle, if an eigenvector satisfies this single returning cone condition, it must satisfy all the others around the full cycle. The returning cone can here be given as a function of $(\kappa_1, \kappa_2)$ without generating overly complicated expressions:
\begin{equation*}
C(\kappa_1, \kappa_2)=\{ v | Rv >0\} = \left\{ v \left| \frac{2 (v_6 - v_7 - 2 \kappa_1 v_6 + 2 \kappa_1 v_5 + 2 \kappa_2 v_5 - 4 \kappa_1 \kappa_2 v_5)}{1 - 2 \kappa_1} > 0 \right.\right\}.
\end{equation*}

\section*{Appendix B}
The matrix $B_P$ can be represented as $D_1CD_2$ where
$D_1=\text{diag}(1,1,d_2,d_1,d_2,1,d_2,d_1,d_2)$, $D_2=\text{diag}(1,\frac{1}{d_1},\frac{1}{d_2},\frac{2\kappa_1}{d_1},1,\frac{1}{d_1},\frac{1}{d_2},\frac{1}{d_1},\frac{1}{d_2})$, and
\begin{equation*}
C=
\begin{pmatrix} 
0&  0&  0& \frac{1}{2\kappa_1}&           0& -1&  0& 0&  0\\
0&  0&  0& \frac{(1-d_1d_2)}{2\kappa_1}&  1& -1&  0& 0&  0\\
0&  0&  0& 1&                             0& -1&  1& 0&  0\\
0&  0&  0& 1&                             0& -1&  0& 1&  0\\
0&  0&  0& 1&                             0& -1&  0& 0&  1\\
1&  0&  0& 1&                             0& -1&  0& 0&  0\\
0&  1&  0& 1&                             0& -1&  0& 0&  0\\
0&  0&  0& 1&                             0& -1&  0& 0&  0\\
0&  0&  1& 1&                             0& -1&  0& 0&  0
\end{pmatrix}.
\end{equation*}
Then eigenvalues of $B_P=D_1 C D_2$ are also eigenvalues of $C D_2 D_1$ since if $B_Pw=\lambda w$ then $C D_2 D_1(CD_2 w)=C D_2 B_P w=\lambda (CD_2 w)$. The characteristic equation for $C D_2 D_1$ is given by
\begin{equation*}
    |C D_2 D_1 - \lambda I|=\left| \begin{array}{ccccccccc}
    -\lambda & 0 & 0 & 1 & 0 & -\frac{1}{d_1} & 0 & 0 & 0 \\
    0 & -\lambda & 0 & 1-d_1d_2 & d_2 & -\frac{1}{d_1} & 0 & 0 & 0 \\
    0 & 0 & -\lambda & 2\kappa_1 & 0 & -\frac{1}{d_1} & 1 & 0 & 0 \\
    0 & 0 & 0 & 2\kappa_1-\lambda & 0 & -\frac{1}{d_1} & 0 & 1 & 0 \\
    0 & 0 & 0 & 2\kappa_1 & -\lambda & -\frac{1}{d_1} & 0 & 0 & 1 \\
    1 & 0 & 0 & 2\kappa_1 & 0 & -\frac{1}{d_1}-\lambda & 0 & 0 & 0 \\
    0 & \frac{1}{d_2} & 0 & 2\kappa_1 & 0 & -\frac{1}{d_1} & -\lambda & 0 & 0 \\
    0 & 0 & 0 & 2\kappa_1 & 0 & -\frac{1}{d_1} & 0 & -\lambda & 0 \\
    0 & 0 & 1 & 2\kappa_1 & 0 & -\frac{1}{d_1} & 0 & 0 & -\lambda \\
    \end{array} \right| =0,
\end{equation*}
which, expanding about the fifth column, and then about the second column in each of the two resulting terms, becomes
\begin{eqnarray*}
    |C D_2 D_1 - \lambda I| =& -\left| \begin{array}{ccccccc}
    -\lambda & 0 & 1  & -\frac{1}{d_1} & 0 & 0 & 0 \\
    0 & -\lambda & 2\kappa_1 & -\frac{1}{d_1} & 1 & 0 & 0 \\
    0 & 0 & 2\kappa_1-\lambda & -\frac{1}{d_1} & 0 & 1 & 0 \\
    0 & 0 & 2\kappa_1 & -\frac{1}{d_1} & 0 & 0 & 1 \\
    1 & 0 & 2\kappa_1  & -\frac{1}{d_1}-\lambda & 0 & 0 & 0 \\
    0 & 0 & 2\kappa_1  & -\frac{1}{d_1} & 0 & -\lambda & 0 \\
    0 & 1 & 2\kappa_1  & -\frac{1}{d_1} & 0 & 0 & -\lambda \\
    \end{array} \right| \\
    & +\lambda^2 \left| \begin{array}{ccccccc}
    -\lambda & 0 & 1  & -\frac{1}{d_1} & 0 & 0 & 0 \\
    0  & -\lambda & 2\kappa_1 & -\frac{1}{d_1} & 1 & 0 & 0 \\
    0 & 0 & 2\kappa_1-\lambda & -\frac{1}{d_1} & 0 & 1 & 0 \\
    1 & 0 & 2\kappa_1  & -\frac{1}{d_1}-\lambda & 0 & 0 & 0 \\
    0 & 0 & 2\kappa_1  & -\frac{1}{d_1} & -\lambda & 0 & 0 \\
    0 & 0 & 2\kappa_1  & -\frac{1}{d_1} & 0 & -\lambda & 0 \\
    0 & 1 & 2\kappa_1  & -\frac{1}{d_1} & 0 & 0 & -\lambda \\
    \end{array} \right| \\
    & -\frac{\lambda}{d_2} \left| \begin{array}{ccccccc}
    -\lambda & 0 & 1  & -\frac{1}{d_1} & 0 & 0 & 0 \\
    0 & 0 & 1-d_1d_2 & -\frac{1}{d_1} & 0 & 0 & 0 \\
    0 & -\lambda & 2\kappa_1 & -\frac{1}{d_1} & 1 & 0 & 0 \\
    0 & 0 & 2\kappa_1-\lambda & -\frac{1}{d_1} & 0 & 1 & 0 \\
    1 & 0 & 2\kappa_1  & -\frac{1}{d_1}-\lambda & 0 & 0 & 0 \\
    0 & 0 & 2\kappa_1  & -\frac{1}{d_1} & 0 & -\lambda & 0 \\
    0 & 1 & 2\kappa_1  & -\frac{1}{d_1} & 0 & 0 & -\lambda \\
    \end{array} \right| = 0,
\end{eqnarray*}
where the final determinant evaluates to $0$, since there is a linear dependence between the second, fifth, and seventh columns. Thus, eigenvalues of $CD_2D_1$ and therefore of $B_P$ are independent of $d_2$ and, consequently, independent of $\kappa_2$. $\qed$

\bibliographystyle{plain} 
\bibliography{trng}
\end{document}